\documentclass[a4paper,twocolumn,showpacs,preprintnumbers,amsmath,amssymb,floatfix]{revtex4}
\usepackage[dvips]{graphicx}
\usepackage{dcolumn}
\usepackage{bm}

\usepackage{literat}
\usepackage{geometry}
\usepackage{amsmath}
\newcommand{\ve}{\varepsilon}

\newcommand{\D}{{\rm d}}
\begin{document}
\title{Phase transitions induced by noise cross--correlations}
\author{A.I.~Olemskoi, D.O.~Kharchenko, I.A.~Knyaz'}
\email{alex@ufn.ru} 
\date{\today}
\begin{abstract}
A general approach to consider spatially extended stochastic
systems with correlations between additive and multiplicative
noises subject to nonlinear damping is developed. Within modified
cumulant expansion method, we derive an effective Fokker--Planck
equation whose stationary solutions describe a character of
ordered state. We find that fluctuation cross--correlations lead to
a symmetry breaking of the distribution function even in the
case of the zero--dimensional system. In general case, continuous,
discontinuous and reentrant noise induced phase transitions take
place. It is appeared the cross--correlations play a role of bias
field which can induce a chain of phase transitions being
different in nature. Within mean field approach, we give an
intuitive explanation of the system behavior through an effective
potential of thermodynamic type. This potential is written in the
form of an expansion with coefficients defined by temperature,
intensity of spatial coupling, auto- and cross--correlation times
and intensities of both additive and multiplicative noises.
\end{abstract}

\pacs{05.10.Gg, 05.40.Ca, 05.70.Fh}
\maketitle
\section{Introduction}
A considerable activity concerning the stochastic processes addresses the
constructive role of fluctuations of environment being called as noises. An
incomplete list of such processes includes noise induced unimodal--bimodal
transitions \cite{Horst}, stochastic resonance \cite{WM95}, noise induced
spatial patterns and phase transitions \cite{Garcia}, etc. In
general case, considering a stochastic dynamics, one should deal with a problem
to account for correlations between random sources.
Along this line several special methods were developed.
Most popular of them are as follows:
(i) the cumulant expansion method \cite{VanKampen,Shapiro};
(ii) the spectral width expansion method \cite{Horst,Risken};
(iii) the unified colored noise approximation where evolution equations
for both a stochastic variable and a random force are combined within
unique equation of motion \cite{UCNA,Mangioni,EPJ2002,PhysA2002}.

The wide spectra of works are aimed to explore an effect of
correlations of fluctuations in extended systems (see
Ref.\cite{Garcia} and references therein). It is appeared the
interplay among noise correlations, nonlinearity and spatial
coupling leads to a special type of noise induced phase transition
known as reentrant transition \cite{VB94}. Moreover, such
reentrance can be observed even in the limit of weakly correlated
noise: if the noise self--correlation time $\tau\to 0$, the system
undergoes a single reentrant phase transition \cite{VB97}.
In the opposite limit of the strong correlated noise ($\tau\to
\infty$) a chain of reentrant phase transitions can take place
\cite{Mangioni}. A kind of these transitions is continuous or
discontinuous in dependence of the self--action part of a bare
potential.

Quite peculiar picture is observed in the case of  several noises
whose cross--correlations can arrive at the remarkable and
counterintuitive phenomena related to the problem of
reconstruction of phase transitions. In such a case the stochastic
system undergoes a chain of different phase transitions with
appearance of a metastable phase in spite of the fact that bare
potential does not assume such a phase \cite{EPJB2003}.
Unfortunately, nowadays we have a scanty collection of works
devoted to this phenomenon. Perhaps, it can be explained by
problems in realization of natural or computer experiments in
extended systems, on the one hand, and by absence of theoretical
tools and methods to perform the corresponding calculations, on
the other hand.

Above pointed phenomena force to reconsider before developed ideas concerning
the theoretical approaches of noise induced phase transitions in extended
systems. Most articles concern the problem of above
cross--correlations within framework of special models which are difficult
to generalize for description of reconstruction of phase transitions.
Therefore, the fluctuation induced rebuilding in the system behavior is an
open question that should be developed to find the unique role of stochastic
environment influence.

In this paper, we consider the general situation to present the
role of cross-correlation contribution of two noises into the
picture of phase transitions. We explore an extended stochastic
system which obeys the archetypal model of
Brownian particle. The adequate scheme which allows to specify
statistical properties of the system with nonlinear kinetic
coefficient in the overdamped limit is introduced. Within a
simplest model with nonlinear damping, drift caused by
Landau--like potential and two colored (multiplicative
and additive) noises, we show in what a way the system can undergo
noise induced phase transitions. We find out  phase transitions
of both continuous and discontinuous character are realized as
biased phase transitions. We obtain phase and bifurcation diagrams
to elucidate the cross--correlation is one of reasons to appear
the metastable states inherent in first order phase transitions.

The paper is organized in the following manner. Section II is devoted to development of
analytical approach to study noise induced phase transitions on the basis of
kinetic equation for the probability density function and cumulant expansion
method.
In Section III, we apply the derived formalism to consider noise
induced phase transitions in the simplest case of the
Ginzburg--Landau model with both multiplicative and additive
noises and kinetic coefficient dependent of the stochastic
variable. On the basis of obtained drift and diffusion
coefficients we build out an effective one--dimensional stochastic
process being multiplicative but noncorrelated. Related
probability distribution function combined with self--consistency
condition allow us to investigate the corresponding phase diagram
and stationary behavior of the order parameter (Section IV).
Discussion in Section V is based on representation of the
effective stochastic process within framework of the mean field
approach. Such consideration allows us to study noise induced
phase transitions by analogy with the standard Landau scheme. It
is appeared the noise cross--correlations and strengthening
dispersion of damping coefficient arrive at transformation of the
continuous phase transition into discontinuous one. Finally, main
results and perspectives of the work are collected in Conclusion
(Section VI).

\section{Main relations} \label{sec:s1}

Consider a Brownian particle under influence of an effective
potential $\mathcal{F}[x(\mathbf{r})]$ and a damping characterized
by the viscosity coefficient $\gamma(x)$. The generalized
evolution equations for the scalar variable $x$ and the conjugate
momentum $p$ read:
\begin{eqnarray}
&&m\dot x=p,
\label{1}\\
&&\dot p +\gamma(x)p= -\frac{\delta \mathcal{F}}{\delta
x(\mathbf{r},t)}+g_\mu(x)\zeta_\mu(\mathbf{r},t)\label{1a}
\end{eqnarray}
where $m$ is the effective particle mass, dot stands for the derivative with
respect to the time $t$, $\mathbf{r}$ is the space coordinate; the effective
potential is reduced to the Ginzburg--Landau form
\begin{equation}
\mathcal{F}=\int \left[V_0(x)+\frac{D}{2}|\nabla x|^2 \right]
\D\mathbf{r}
\end{equation}
with $V_0(x)$ and $D>0$ being a specific thermodynamic potential and
an inhomogeneity constant, $\nabla \equiv
\partial/\partial \mathbf{r}$. The last term in Eq.(\ref{1a}), where
index $\mu=1,2$ numerates different noises to be summarized in accordance with
the Einstein rule, represents Langevin forces which act with amplitudes
$g_\mu(x)$ and stochastically alternating functions $\zeta_\mu$. Neglecting a
space correlation, we focus on time correlations between forces $\zeta_\mu$,
i.e.,
\begin{equation}
\langle\zeta_\mu(\mathbf{r},t)\zeta_\nu(\mathbf{r}',t')\rangle
=\delta(\mathbf{r}-\mathbf{r'})C_{\mu\nu}(t-t').
\end{equation}
Inserting Eq.(\ref{1a}) into the result of differentiating
Eq.(\ref{1}) over time, it is convenient to represent
evolution equation for the quantity $x$ in the form
\begin{equation}
m\ddot x+\gamma(x)\dot x= f(x)+D\Delta x+g_\mu(x)\zeta_\mu
\label{2a}
\end{equation}
where $f(x)=-\D V_0/\D x$ is a deterministic force.

To study statistical properties of the system one needs to find
the probability density function $P=P(p,x,t)$ of the system states
distribution in the phase state $\{x,p\}$. With this aim, we
represent the system on the regular $d$--dimension lattice with
mesh size $\Delta \ell=1$ and $\gamma_i=\gamma(x_i)$,
$f_i=f(x_i)$, $g_{\mu i}=g_\mu(x_i)$. Then, the differential
equation with partial derivatives (\ref{2a}) is reduced to the
usual differential equation
\begin{equation}\label{2}
\ddot x_i+\gamma_i\dot
x_i=f_i+\frac{D}{2d}\sum_{j}\widehat{D}_{ij}x_j+g_{\mu
i}\zeta_{\mu i}
\end{equation}
due to representation of the Laplacian operator on a grid as
follows
\begin{equation} \Delta\rightarrow\Delta_i\equiv\sum_j
\widehat{D}_{ij}=\sum_j(\delta_{nn(i),j}-2d\delta_{ij})
\end{equation}
where $nn(i)$ notices a set of the nearest neighbors of the site $i$.

By definition, the probability density function is given by
averaging over noises of the density function $\rho(x_i,p_i,t)$ of
the microscopic states distribution in the phase space:
\begin{equation}
P(x_i, p_i,t)=\langle\rho(x_i, p_i,t)\rangle.\label{X}
\end{equation}
To construct an equation for the macroscopic density function
$P=P(x_i, p_i,t)$ we exploit the conventional device to proceed
from the continuity equation for the microscopic one
$\rho=\rho(x_i,p_i,t)$:
\begin{equation}\label{ce6}
\frac{\partial\rho}{\partial t}+\left[\frac{\partial}{\partial
x_i}(\dot x_i\rho)+\frac{\partial}{\partial p_i}(\dot
p_i\rho)\right]=0.
\end{equation}
Inserting the time derivative of the momentum $\dot p=m\ddot x$
from Eq.(\ref{2}) into Eq.(\ref{ce6}), we obtain
\begin{equation}\label{ce7}
\frac{\partial\rho}{\partial t}= \left(\widehat{\mathcal{L}}+
\widehat{\mathcal{N}}_{\mu}\zeta_{\mu}\right)\rho
\end{equation}
where the operators $\widehat{\mathcal{L}}$ and
$\widehat{\mathcal{N}}_{\mu}$ are defined as follows:
\begin{eqnarray}
 &&\widehat{\mathcal{L}}\equiv-\frac{p_i}{m}\frac{\partial}{\partial x_i}\nonumber\\&&-\frac{\partial}{\partial p_i}\left(f_i+\frac{D}{2d}\sum_j \widehat{D}_{ij}
x_j-\frac{\gamma_i}{m}p_i \right),\label{AA}\\
&&\widehat{\mathcal{N}}_{\mu}\equiv -g_{\mu i} \frac{\partial}{\partial
p_i}\label{BB}.
\end{eqnarray}
Within the interaction representation, the microstate density
function reads as
\begin{equation}
\wp=e^{-\widehat{\mathcal{L}}t}\rho
\end{equation} to reduce
Eq.(\ref{ce7}) to the form
\begin{eqnarray}
&\frac{\partial\wp}{\partial t}=\sum\limits_{\mu}
\widehat{\mathcal{R}}_{\mu}\wp,\label{a}\\
&\widehat{\mathcal{R}}_{\mu}=
\widehat{\mathcal{R}}_{\mu}(x_i,p_i,t)\equiv\zeta_{\mu}\left(
e^{-\widehat{\mathcal{L}}t}\widehat{\mathcal{N}}_{\mu}
e^{\widehat{\mathcal{L}}t}\right).\label{b}
\end{eqnarray}
A standard and effective device to solve such a type of stochastic equation is
the well--known cumulant expansion method \cite{VanKampen}. Neglecting terms of
the order $O(\widehat{\mathcal{R}}_{\mu}^3)$, we arrive at the kinetic equation
of the form
\begin{widetext}
\begin{equation}\label{ce12}
\begin{split}
\frac{\partial}{\partial t} \langle\wp\rangle(t)=\left[
\sum_{\mu\nu}\int\limits_0^t\langle\widehat{\mathcal{R}}_{\mu}(t)
\widehat{\mathcal{R}}_{\nu}(t')\rangle\D
t'\right]\langle\wp\rangle(t).
\end{split}
\end{equation}
Within the original representation, the  equation for the probability
density (\ref{X}) reads
\begin{equation}
\begin{split}
\frac{\partial}{\partial t}P(t)=\left\{\widehat{\mathcal{L}}+
\int\limits_0^tC_{\mu\nu}(\tau)\left[\widehat{\mathcal{N}}_{\mu}\left(
e^{\widehat{\mathcal{L}}\tau}\widehat{\mathcal{N}}_{\nu}e^{-\widehat{\mathcal{L}}\tau}
\right)\right]\D\tau\right\}P(t).
\end{split}
\end{equation}
\end{widetext}
If the physical time is much more than a
correlation scale $(t\gg \tau_\mu)$, we can replace the upper limit of
the integration by $t=\infty$. Then, expanding exponents, we derive
to the perturbation expansion
\begin{equation}
\frac{\partial P}{\partial
t}=\left(\widehat{\mathcal{L}}+\widehat{\mathcal{C}}\right)P
\label{FP}
\end{equation}
where collision operator
\begin{equation}
\widehat{\mathcal{C}}\equiv\sum_{n=0}^\infty\widehat{\mathcal{C}}^{(n)},\quad
\widehat{\mathcal{C}}^{(n)}\equiv
M_{\mu\nu}^{(n)}\left(\widehat{\mathcal{N}}_{\mu}
\widehat{\mathcal{L}}^{(n)}_{\nu}\right)\label{a1}
\end{equation}
is determined through the commutators
\begin{equation}
\widehat{\mathcal{L}}_{\nu}^{(n+1)}=[\widehat{\mathcal{L}},\widehat{\mathcal{L}}_{\nu}^{(n)}],
\quad\widehat{\mathcal{L}}_{\nu}^{(0)}\equiv\widehat{\mathcal{N}}_{\nu}
\end{equation}
and moments of correlation function
\begin{equation}
M_{\mu\nu}^{(n)}=\frac{1}{n!}\int_0^\infty
\tau^nC_{\mu\nu}(\tau){\rm d}\tau.\label{b1}
\end{equation}

To perform next calculations we shall restrict ourselves considering overdamped
systems where the variation scales $t_s$, $x_s$, $v_s$, $\gamma_s$, $f_s$, and
$g_s$ of the time $t$, the quantity $x$, the velocity $v\equiv p/m$, the
damping coefficient $\gamma(x)$, the force $f(x)$ and the noise amplitudes
$g_{\mu}(x)$, respectively, obey the following conditions:
\begin{equation}\label{scale}
\begin{split}
\frac{v_s t_s}{x_s}\equiv\epsilon^{-1}\gg 1,\quad \gamma_s t_s=\epsilon^{-2}\ggg 1,\\
\frac{f_s t_s}{v_s m}=\epsilon^{-1}\gg 1,\quad\frac{g_s t_s}{v_s
m}=\epsilon^{-1}\gg 1.
\end{split}
\end{equation}
These conditions means a hierarchy of the damping and the
deterministic/stochastic forces to be characterized by relations
\begin{equation}\label{scale1}
\frac{f_s/m}{\gamma_s v_s}=\epsilon\ll
1,\quad\frac{g_s/m}{\gamma_s v_s}=\epsilon\ll 1.
\end{equation}

As a result, the dimensionless system of equations (\ref{1}), (\ref{1a}) takes
the form
\begin{equation}\label{eq22}
\begin{split}
\frac{\partial x}{\partial t}=&~\epsilon^{-1}v,\\
\frac{\partial v}{\partial t}=&-\epsilon^{-2}\gamma(x)v\\
&+\epsilon^{-1}\left[f(x)+D\Delta
x+g_\mu(x)\zeta_\mu(\mathbf{r},t)\right].
\end{split}
\end{equation}
\vspace{0.1mm}

\noindent Respectively, the Fokker--Planck equation (\ref{FP}) reads
\begin{eqnarray}\label{Ptheta}
\left(\frac{\partial }{\partial
t}-\widehat{\mathcal{L}}\right)P=\epsilon^{-2}\widehat{\mathcal{C}}P
\end{eqnarray}
where the operator
\begin{equation}
\widehat{\mathcal{L}}\equiv\epsilon^{-1}\widehat{\mathcal{L}}_1
+\epsilon^{-2}\widehat{\mathcal{L}}_2
\end{equation}
has the components
\begin{equation}
\begin{split}
&\widehat{\mathcal{L}}_1\equiv -v_i\frac{\partial}{\partial x_i}
-\left(f_i+\frac{D}{2d}\sum_{j}\widehat{D}_{ij}x_j\right)\frac{\partial}{\partial
v_i},\\
&\widehat{\mathcal{L}}_2\equiv\gamma_i~\frac{\partial}{\partial
v_i}v_i.
\end{split}
\end{equation}
The collision operator $\widehat{\mathcal{C}}$ is defined through the
expressions (\ref{a1}) --- (\ref{b1}) and the operator (\ref{BB}) with the
momentum $p_i$ being replaced by the velocity $v_i$. With accuracy up to the
first order in $\epsilon\ll 1$ the expansion (\ref{a1})
takes the explicit form
\begin{widetext}\label{scat}
\begin{equation}
\widehat{\mathcal{C}}=\left(M^{(0)}_{\mu\nu}-\gamma_i
M^{(1)}_{\mu\nu}\right) g_{\mu i}g_{\nu
i}\frac{\partial^2}{\partial v_i^2}+\epsilon
M^{(1)}_{\mu\nu}g_{\mu i}g_{\nu i}\left[-\frac{1}{g_{\nu
i}}\left(\frac{\partial g_{\nu i}}{\partial
x_i}\right)\left(\frac{\partial}{\partial v_i}
+v_i\frac{\partial^2}{\partial
 v_i^2}\right)+\frac{\partial^2}{\partial
 x_i\partial
 v_i}\right]+O(\epsilon^2).
\end{equation}
\end{widetext}

To obtain the usual probability function ${\mathcal P}(x_i,t)$ we
consider velocity moments of the initial distribution function
$P(x_i,v_i,t)$ in the standard form \cite{Risken}
\begin{equation}\label{one_site}
{\mathcal P}_n(x_i,t)\equiv\int v_i^n P(x_i,v_i,t){\rm d}v_i
\end{equation}
where integration over all set $\{v_i\}$ is performed. Then, if we multiply
the Fokker--Planck equation (\ref{Ptheta}) by the factor
$v^n$ and integrate over velocities, we derive
the following recurrent relations:
\begin{widetext}
\begin{equation}\label{Pmoment}
\begin{split}
&\epsilon^2\frac{\partial{\mathcal P}_n}{\partial
t}+n\gamma_i{\mathcal P}_n+\epsilon\left[\frac{\partial{\mathcal
P}_{n+1}}{\partial
x_i}-n\left(f_i+\frac{D}{2d}\sum_{j}\widehat{D}_{ij}x_j\right)
{\mathcal P}_{n-1}\right]\\
&=n(n-1)\left(M^{(0)}_{\mu\nu}-\gamma_i M^{(1)}_{\mu\nu}
\right)g_{\mu i}g_{\nu i}{\mathcal P}_{n-2}-\epsilon n
M^{(1)}_{\mu\nu}\left[g_{\mu i}g_{\nu i}\frac{\partial{\mathcal
P}_{n-1}}{\partial x_i}+n g_{\mu i}\left(\frac{\partial g_{\nu
i}}{\partial x_i}\right){\mathcal P}_{n-1}\right]+O(\epsilon^2).
\end{split}
\end{equation}
\end{widetext}

At $n=0$, we obtain the equation for the distribution function
${\mathcal P}\equiv{\mathcal P}_0(x_i,t)$:
\begin{equation}
\frac{\partial{\mathcal P}}{\partial
t}=-\epsilon^{-1}\frac{\partial{\mathcal P}_1}{\partial x_i}.
\end{equation}
The expression for the first moment ${\mathcal P}_1$ follows from
Eq.(\ref{Pmoment}) where $n=1$ and only terms of the first order in $\epsilon$
are kept:
\begin{equation}
\begin{split}
{\mathcal P}_1=\frac{\epsilon}{\gamma_i}\left\{\left(f_i+
\frac{D}{2d}\sum_{j}\widehat{D}_{ij}x_j\right){\mathcal
P}\qquad\qquad\right.
\\ \left.
-\frac{\partial{\mathcal P}_2}{\partial
x_i}-M^{(1)}_{\mu\nu}\left[g_{\mu i}g_{\nu
i}\frac{\partial{\mathcal P}}{\partial x_i}+g_{\mu
i}\left(\frac{\partial g_{\nu i}}{\partial x_i}\right){\mathcal
P}\right]\right\}.
\end{split}
\end{equation}
The second moment ${\mathcal P}_2$ can be obtained if one puts in
Eq.(\ref{Pmoment}) $n=2$ and takes into account only zeroth terms of smallness
over the parameter $\epsilon\ll 1$:
\begin{equation}
{\mathcal P}_2=\left(\frac{M^{(0)}_{\mu\nu}}{\gamma_i}-
M^{(1)}_{\mu\nu}\right)g_{\mu i}g_{\nu i}{\mathcal P}.
\end{equation}
At last, the Fokker--Planck equation takes the Kramers--Moyal form
\begin{equation} \label{Fok_Plank}
\frac{\partial{\mathcal P}}{\partial t}=-\frac{\partial }{\partial
x_i}\left(\mathcal{D}_1{\mathcal P}\right)+\frac{\partial^2
}{\partial x^2_i}\left(\mathcal{D}_2{\mathcal P}\right),
\end{equation}
where effective drift and diffusion coefficients are as follows:
\begin{equation}\label{D1}
\begin{split}
&\mathcal{D}_1=\frac{1}{\gamma_i}\left\{\left(f_i+\frac{D}{2d}\sum_{j}
\widehat{D}_{ij}x_j\right)\right.\\
&\left.+ \left[M^{(0)}_{\mu\nu}g_{\mu i}g_{\nu i}\frac{\partial
\gamma^{-1}_i}{\partial x_i}+M^{(1)}_{\mu\nu}g_{\mu
i}\left(\frac{\partial g_{\nu i}}{\partial
 x_i}\right)\right]\right\},
\end{split}
\end{equation}
\begin{equation}\label{D2}
\mathcal{D}_2=\frac{M^{(0)}_{\mu\nu}}{\gamma_i^2}~g_{\mu i}g_{\nu
i}.
\end{equation}

To proceed the consideration we need to pass from the grid representation to a
continuous one. In so doing, we use the mean field approximation to replace
the second term of effective interaction force in Eq.(\ref{D1}):
\begin{equation}
\frac{D}{2d}\sum_{j}\widehat{D}_{ij}x_j\to D(\eta-x)
\label{eta}
\end{equation}
where an order parameter $\eta\equiv \langle x\rangle$ is defined through the
self--consistency equation
\begin{equation} \label{selconst}
\eta=\int\limits_{-\infty}^{\infty} x{\mathcal P}_\eta(x){\rm d}x,
\end{equation}
${\mathcal P}_\eta(x)$ is a solution of the Fokker--Planck equation
(\ref{Fok_Plank}). Under stationary condition, the relevant distribution
function has the form
\begin{equation}\label{pdf}
{\mathcal
P}_\eta(x)=\frac{\mathcal{Z}^{-1}_\eta}{\mathcal{D}_2(x)}
\exp\left(\int\limits_{-\infty}^x\frac{\mathcal{D}_1(x',\eta)}{\mathcal{D}_2(x')}\D
x'\right)
\end{equation}
where the partition function
\begin{equation}
\mathcal{Z}_\eta=\int\limits_{-\infty}^{\infty}\frac{\D
x}{\mathcal{D}_2(x)}
\exp\left(\int\limits_{-\infty}^x\frac{\mathcal{D}_1(x',\eta)}{\mathcal{D}_2(x')}\D
x'\right)
\end{equation}
takes care of normalization condition. The equation (\ref{selconst}) has
solutions within the domain bounded by the Newton--Raphson condition
\begin{equation} \label{New}
\left.\int\limits_{-\infty}^{\infty} x \frac{\partial}{\partial\eta} {\mathcal
P}_\eta(x)\right|_{\eta=0}{\rm d}x=1
\end{equation}
obtained by differentiating Eq.(\ref{selconst}) over the order parameter
$\eta$.

\section{Model of correlation between additive and multiplicative noises}

To apply the general results obtained in Section \ref{sec:s1} we consider
in details the simplest model of two correlated noises being additive and multiplicative
in nature. Relevant amplitudes are defined as follows:
\begin{equation}
g_a(x)=1, \quad g_m(x)={\rm sign}(x)|x|^a
\label{n2}
\end{equation}
where the exponent is defined as $a\in [0,1]$ and
the sign function is introduced to take into account a direction
of the Langevin force.
We will focus on the prototype system
concerning the Ginzburg--Landau model with the potential
\begin{equation}\label{GL}
V_0(x)=-\frac{\ve}{2}x^2+\frac{1}{4}x^4
\end{equation}
where $\ve$ is a parameter being dimensionless temperature counted off a
critical value in negative direction.
In correspondence with the line of consideration \cite{Bray},
we take up the viscosity coefficient in the form
\begin{equation}
\gamma(x)=|x^2-1|^{-\alpha}
\label{n3}
\end{equation}
where positive index $\alpha$ stands to measure the damping weakening near
bare state $x=1$.

Next, we suppose the noises are to be Gaussian distributed, white in space with
zero mean and colored in time according to the correlation matrix
\begin{equation} \label{matrix}
\widehat{\mathbf{C}}(\tau )=\left(\begin{array}{cc}
\frac{\sigma_a^2}{\tau_a}e^{-|\tau|/\tau_a} &
\frac{\sigma_a\sigma_m}{\tau_c}e^{-|\tau|/\tau_c} \\ & \\
\frac{\sigma_m\sigma_a}{\tau_c}e^{-|\tau|/\tau_c} &
\frac{\sigma_m^2}{\tau_m}e^{-|\tau|/\tau_m}
\end{array}\right)
\end{equation}
where $\sigma_a$ and $\sigma_m$ are amplitudes of additive and multiplicative
noises, respectively, $\tau_a$ and $\tau_m$ are corresponding autocorrelation
times, $\tau_c$ is a time of the cross--correlation between the noises. Moments
(\ref{b1}) of both zero and first orders of correlation matrix (\ref{matrix})
are as follows:
\begin{equation}\label{mc}
\begin{split}
&\widehat{\mathbf{M}}^{(0)}=\left(\begin{array}{cc}\sigma_a^2 &
\sigma_a\sigma_m
\\ \sigma_m\sigma_a &\sigma_m^2
\end{array}\right), \\ &\widehat{\mathbf{M}}^{(1)}=\left(\begin{array}{cc}
\tau_a\sigma_a^2 & \tau_c(\sigma_a\sigma_m)
\\ \tau_c(\sigma_m\sigma_a) & \tau_m\sigma_m^2
\end{array}\right).
\end{split}
\end{equation}
Then, the expressions (\ref{D1}), (\ref{D2}) for effective drift
and diffusion coefficients take the form
\begin{widetext}
\begin{equation} \label{D1a}
\begin{split}
\mathcal{D}_1 = &
|x^2-1|^\alpha\left\{\left[D(\eta-x)+x(\varepsilon-x^2)\right]
+a\sigma_m|x|^{a-1}
\left[\sigma_a\tau_c+\sigma_m\tau_m{\rm sign}(x)|x|^a\right]\right\}\\
& +2\alpha~x(x^2-1)^{2\alpha-1}
\left[\sigma_a+\sigma_m{\rm sign}(x)|x|^a\right]^2,
\end{split}
\end{equation}
\end{widetext}
\begin{equation}\label{D2a}
\mathcal{D}_2=(x^2-1)^{2\alpha}\left[\sigma_a+\sigma_m{\rm sign}(x)
|x|^a\right]^2.
\end{equation}
It is worthwhile to notice at $\gamma(x)={\rm const}$ ($\alpha=0$)
the additive noise can not give a contribution to the drift coefficient
(\ref{D1a}) related to the first of moments (\ref{mc}).

\section{Noise Correlation Induced Transitions}

Addressing to the influence of noise cross--correlations on the
system behavior we start with self--consistency equation
(\ref{selconst}) where the stationary distribution (\ref{pdf}) is
given through the drift and diffusion coefficients (\ref{D1a}),
(\ref{D2a}). It is well known, at phase transitions, the symmetry
breaking causes the ordered state corresponding to the solution
$\eta\ne0$ within the interval of available values of stochastic
quantity $x$, while the disordered phase is related to $\eta=0$.
In the absence of the multiplicative noise, the only reason to
break the symmetry of the stochastic distribution is the
interaction force (\ref{eta}) which plays the role of a conjugate
field related to the order parameter $\eta$. The principle feature
of far-off-equilibrium systems with colored noise is that the
symmetry can be restored due to combined effect of both the
multiplicative noise and the system nonlinearity \cite{Mangioni,VB97}.
Therefore, the reentrant phase transition in such
systems is appeared. Our aim is to demonstrate the
course of the phase transition can be crucially changed by means
of cross--correlations.

\begin{figure}[h]
\centering
\includegraphics[width=80mm]{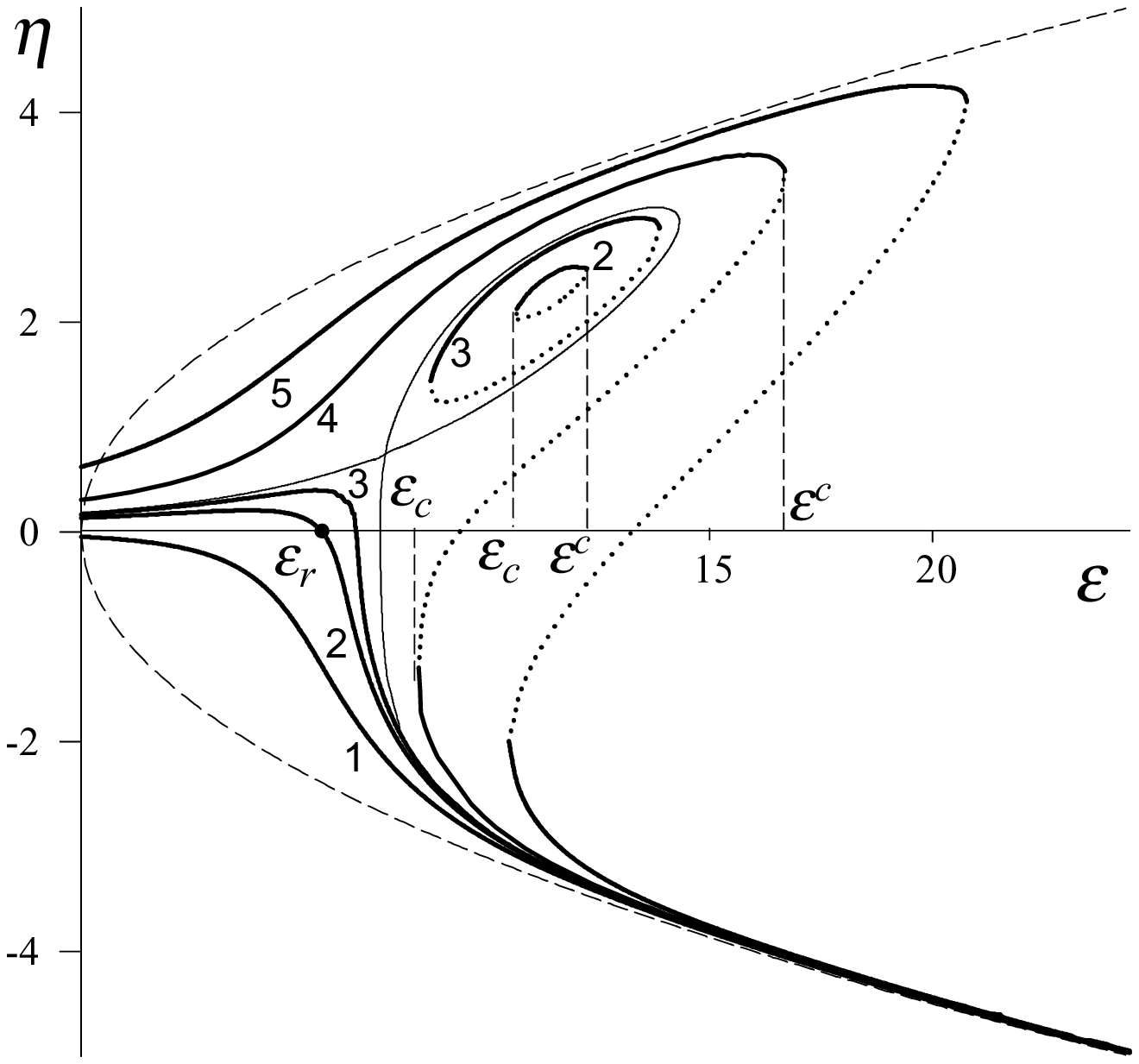}
\caption{Dependence of the order parameter $\eta$ on the
control parameter $\ve$ at
$a=0.8$, $\alpha=0$, $\sigma^2_a=4.84$, $\sigma^2_m=0.01$, $\tau_m=0.01$,
$D=1.0$. Curves 1, 2, 3, 4, 5 correspond to $\tau_c\to 0$,
$\tau_c=2.5$, 3.0, 5.0, 10.0, respectively.
Dashed curve relates to bare dependence $\eta=\pm\sqrt{\ve}$,
dotted curves correspond to unstable solutions.}
\label{eta(e)}
\end{figure}

First, we consider the solution of Eq.(\ref{selconst})
at different values of the noise cross--correlation scale $\tau_c$.
As shown in Figure \ref{eta(e)},
in the absence of both noises ($\sigma_a=\sigma_m=0$) and coupling
($D=0$) the system behaves itself in a usual manner being inherent in the
square--root law (dashed curve with both vertical derivative in
the point of origin and symmetry with respect to the $\ve$--axis).
Such behavior means the maxima appearance of the distribution
(\ref{pdf}) in points $\pm\sqrt{\ve}$ that can be interpreted as
standard noise induced transition of the second order with mean
value $\eta=0$ \cite{UFN}. With switching on noises and coupling
the situation is changed principally. First, the above symmetry is
broken to survive the only negative value of the order parameter
in the limit of small cross--correlations (curve 1). Combined
effect of correlated noises, system nonlinearity and spatial
coupling arrives at the change of the order
parameter sign at small values $\eta$.
Indeed, as seen from curves 2, 3, an increase in the
cross--correlation time $\tau_c$ shifts weakly negative solution into the
positive domain causing the reorientation transition at the driving
parameter $\ve_r$. In addition to this transition, an increase in the
cross--correlation scale arrives at positive solutions appeared according to
discontinuous phase transitions that has reentrant nature
within domain bounded by both lower $\ve_c$ and
upper $\ve^c$ boundaries (see elliptic form parts of curves 2, 3 where solid
and dotted lines relate to (meta)stable \cite{meta} and unstable solutions).
With subsequent growth of the cross--correlation time above the value $\tau_{cr}$
related to thin solid curve in Figure \ref{eta(e)}, back
bifurcation happens and temperature dependence of the order
parameter takes one--connected character. This means an increase in the
control parameter $\ve$ arrives at the (meta)stable branch of positive magnitudes
$\eta$ initially (solid curve), then the unstable branch (dotted
curve) follows from the point $\ve^c$ down to $\ve_c$ and finally
the negative (meta)stable state is merged.
It results in formation of a hysteresis loop in $\eta(\ve)$ dependence
where both (meta)stable and unstable states exist to be
solutions of Eq.(\ref{selconst}) (see curves 4, 5).
Thus, one can conclude both reentrance and reorientation phase transitions
are inherent in systems with colored multiplicative noise.

\begin{figure}[h]
\centering
\includegraphics[width=80mm]{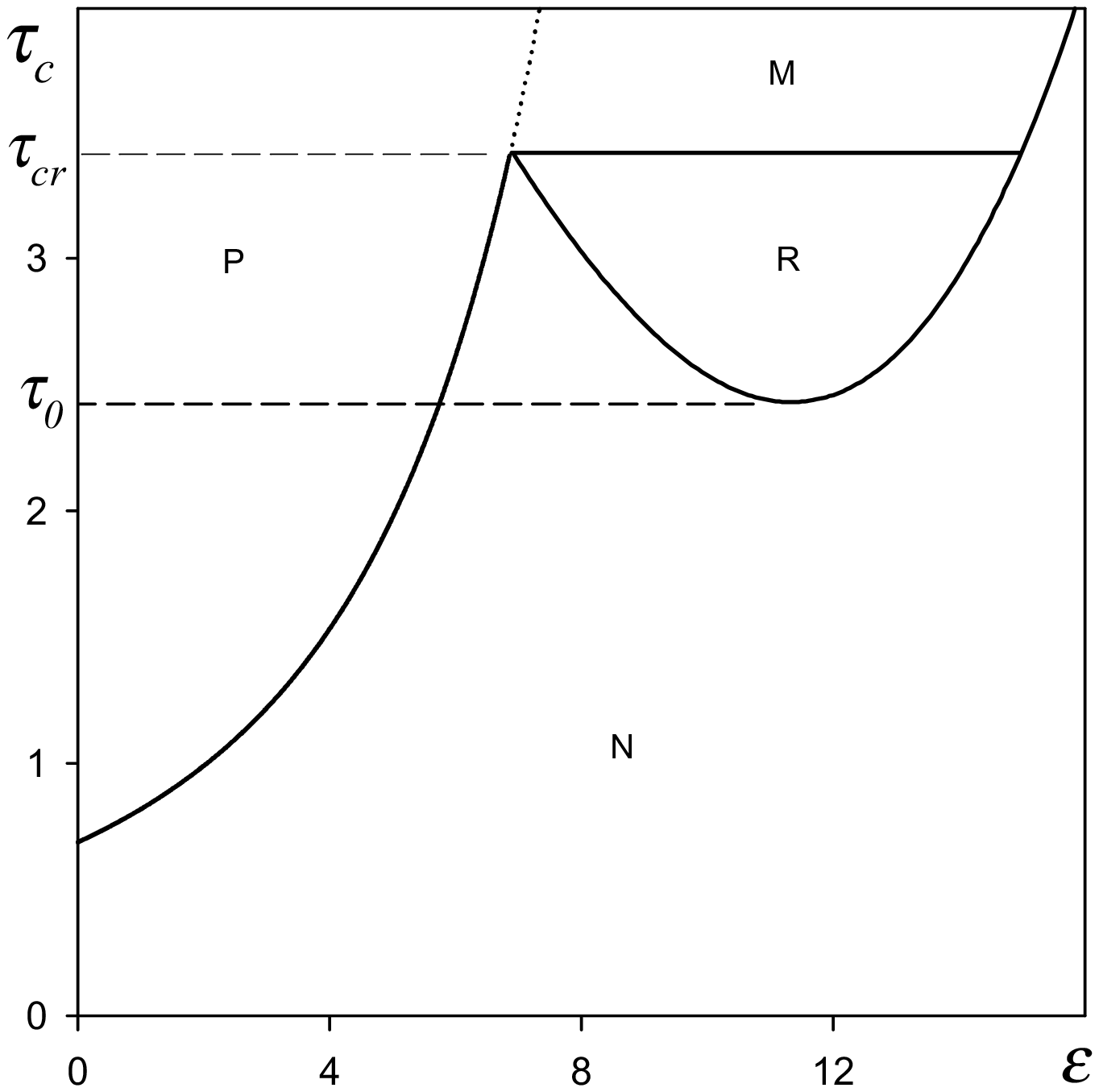}
\caption{Phase diagram in $(\ve,\tau_c)$ plane at $a=0.8$, $\alpha=0$,
$\sigma^2_a=4.84$, $\sigma^2_m=0.01$, $\tau_m=0.01$, $D=1.0$.
Domains denoted as P, N, R and M correspond to positive and negative
$\eta$ values, reentrant transition and (meta)stable phases, respectively.
\label{t(e)}}
\end{figure}

In Figure \ref{t(e)} we plot a phase diagram in ($\ve, \tau_c$) plane
to show the influence of the noise cross--correlation scale on the
bifurcation magnitudes of the control parameter.
It is seen, at small cross--correlation times $\tau_c$, the
negative values of the order parameter $\eta$ is inherent in the
whole $\ve$--domain denoted as N in Figure \ref{t(e)}.
With increase in $\tau_c$ the reorientation phase transition
occurs into the state P related to the positive solutions
$\eta>0$. The line of this transition is determined by the
self--consistency equation (\ref{selconst}) at condition $\eta=0$.
At a magnitude $\tau_0$, a doubly bounded domain R appears to
relate to the reentrant transition.
At the critical value $\tau_{cr}$ corresponding to the bifurcation curve in
Figure \ref{eta(e)} the domain R passes to a region M where
all stable, metastable and unstable phases take place.
Over the critical correlation time
$(\tau_c>\tau_{cr})$, the dotted curve relates to the lower
critical value $\ve_c$ of the control parameter in Figure
\ref{eta(e)} (curve 4).

\begin{figure}[h]
\centering
a\\
\includegraphics[width=80mm]{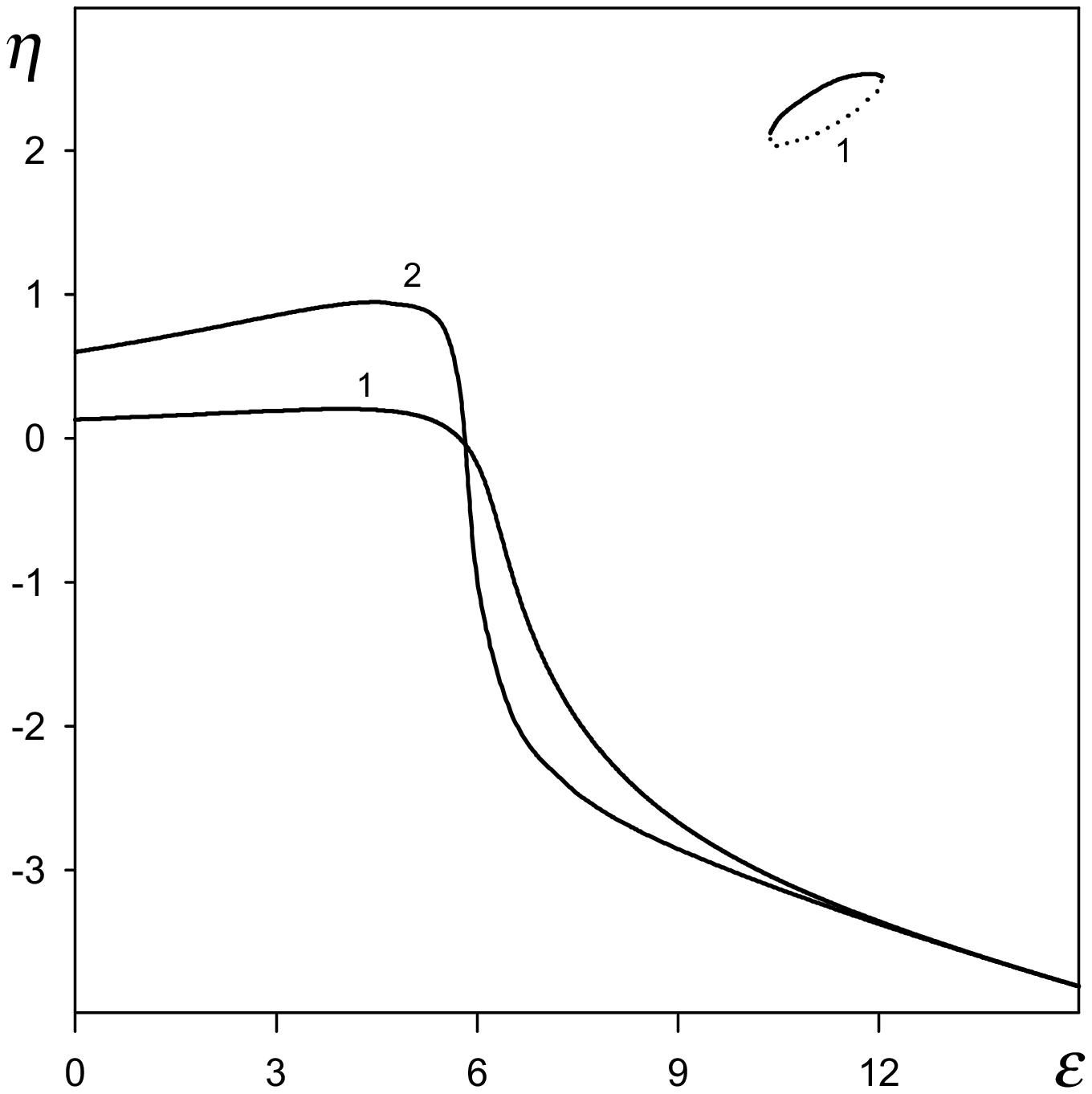}\\
b\\
\includegraphics[width=83mm]{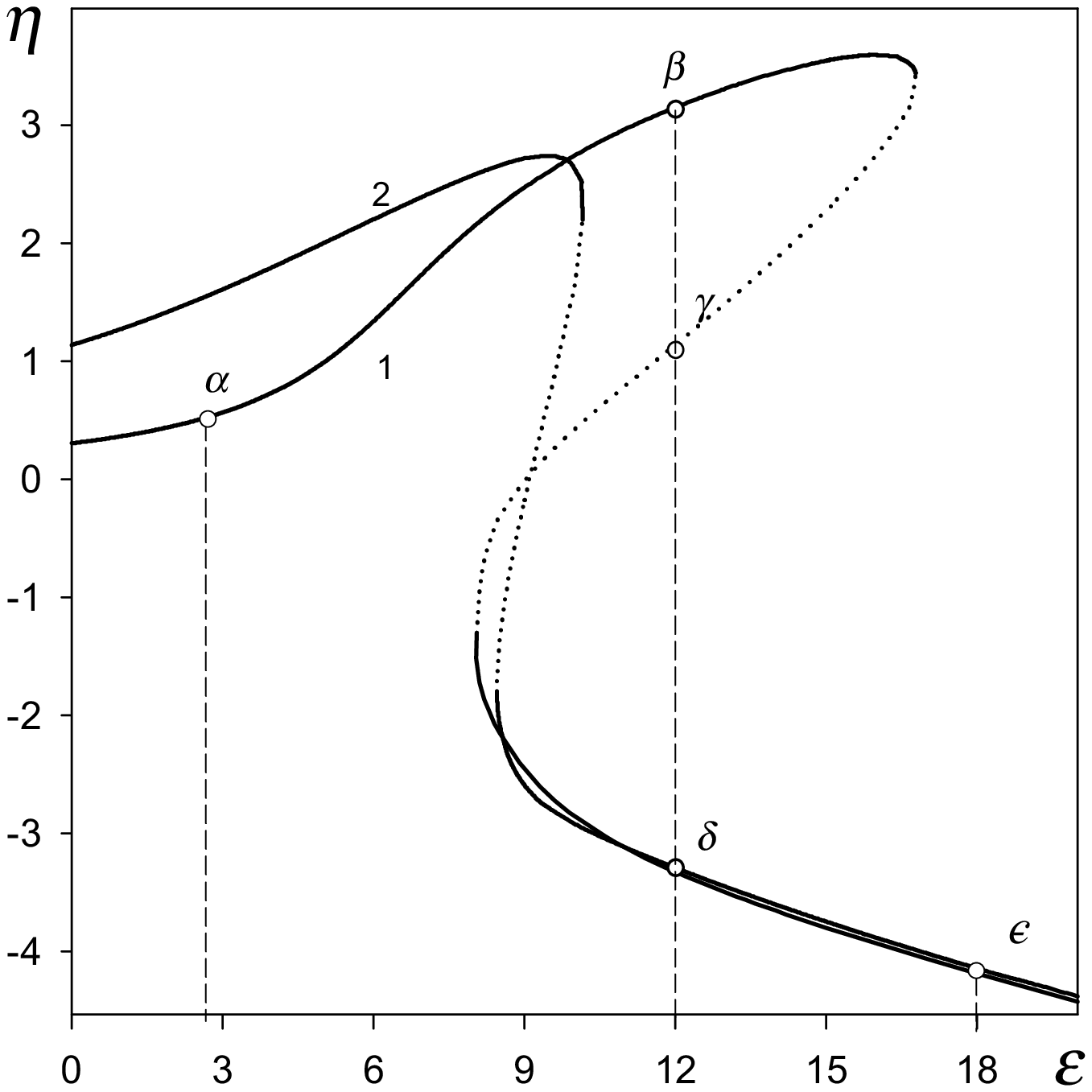}
\caption{The order parameter $\eta$ vs. the control parameter $\ve$ at $a=0.8$,
$\alpha=0.0$, $D=1.0$, $\sigma^2_a=4.84$, $\tau_m=0.01$ and cross--correlation
times $\tau_c=2.5$ (a) and $\tau_c=5.0$ (b).
Curves 1, 2 relates to the multiplicative noise intensities
$\sigma^2_m=0.01$ and  $\sigma^2_m=0.25$.
Points from $\alpha$ to $\epsilon$ address to corresponding curves in Figure
\ref{fig10}.
\label{eta(e)_sigma}}
\end{figure}

The influence of the multiplicative noise intensity $\sigma^2_m$
on the phase transition picture is demonstrated in Figure
\ref{eta(e)_sigma}.
The upper panel relates to moderate
cross--correlation times $\tau_c$ where an increase in $\sigma^2_m$
transforms two--connected $\eta(\ve)$ dependence into
one--connected one varying more fast. With growth of the time
scale $\tau_c$, the increase of the multiplicative noise intensity
arrives at the shrinking the
metastability domain (see lower panel in Figure
\ref{eta(e)_sigma}). Thus, we get the conclusion about dual
role of the multiplicative noise: at small intensities
$\sigma^2_m\ll 1$, main influence is rendered by the
cross--correlations between additive and multiplicative components
of the noise to sharpen the phase transition (see Figure
\ref{eta(e)}); on the other hand, a raising the intensity of the
latter component up to $\sigma^2_m\sim 1$ smears this transition.

\begin{figure}[t]
\centering
\includegraphics[width=80mm]{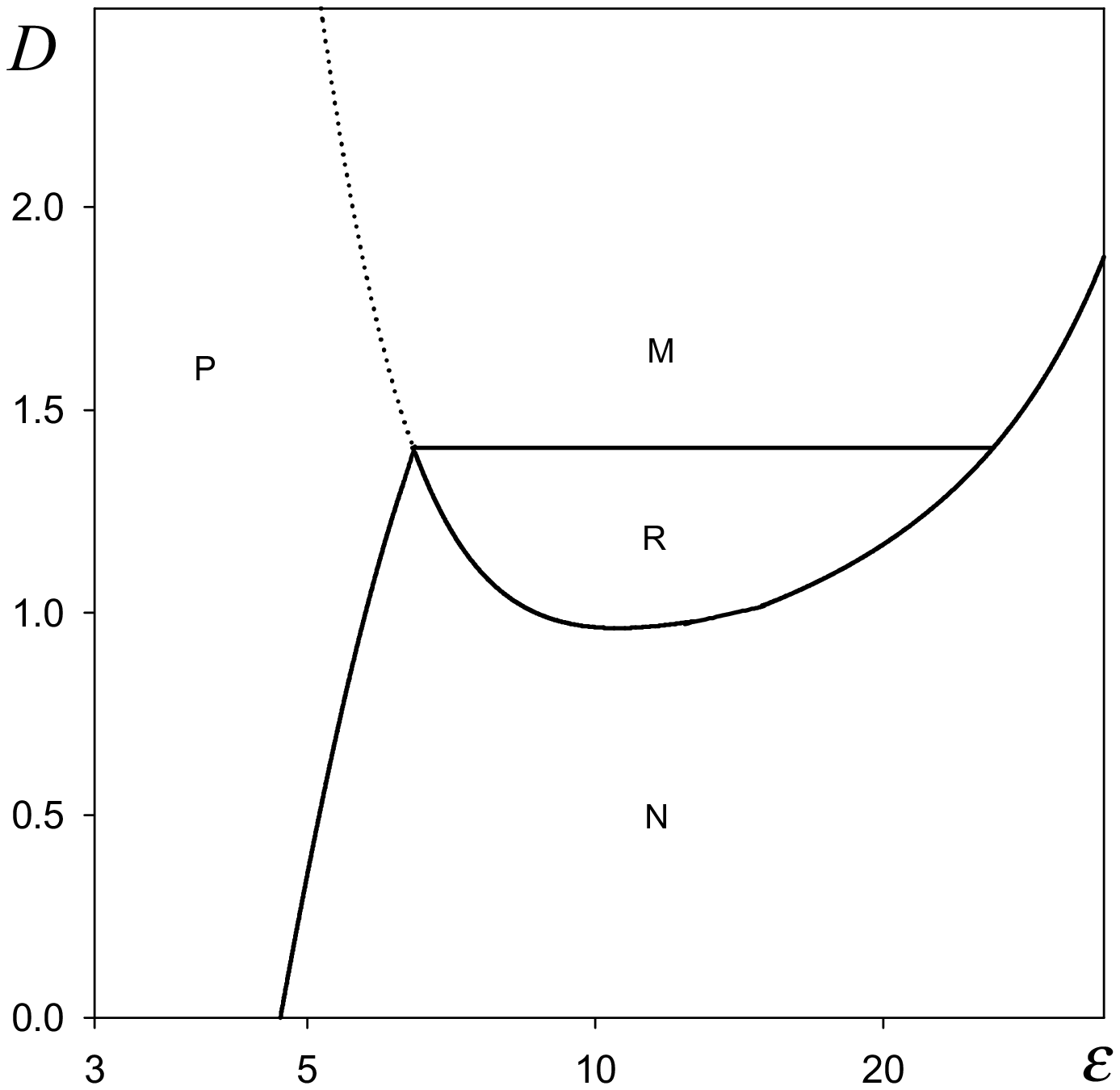}
\caption{Phase diagram in $(\ve, D)$ plane at $a=0.8$, $\alpha=0$,
$\sigma^2_a=4.84$, $\sigma^2_m=0.01$, $\tau_m=0.01$, $\tau_c=2.5$.
\label{e(d)}}
\end{figure}

According to Figure \ref{e(d)} only one difference of the phase
diagram in the axes $(\ve, D)$ from the situation depicted in
Figure \ref{t(e)} takes place. Here, at small magnitudes of the control
parameter $\ve$, the domain P of the positive valued order
parameter $\eta>0$ relates to the whole region of the coupling
parameter $D$ --- contrary to low bounded domain of the
cross--correlation times $\tau_c$.

\begin{figure}[h]
\centering
\includegraphics[width=83mm]{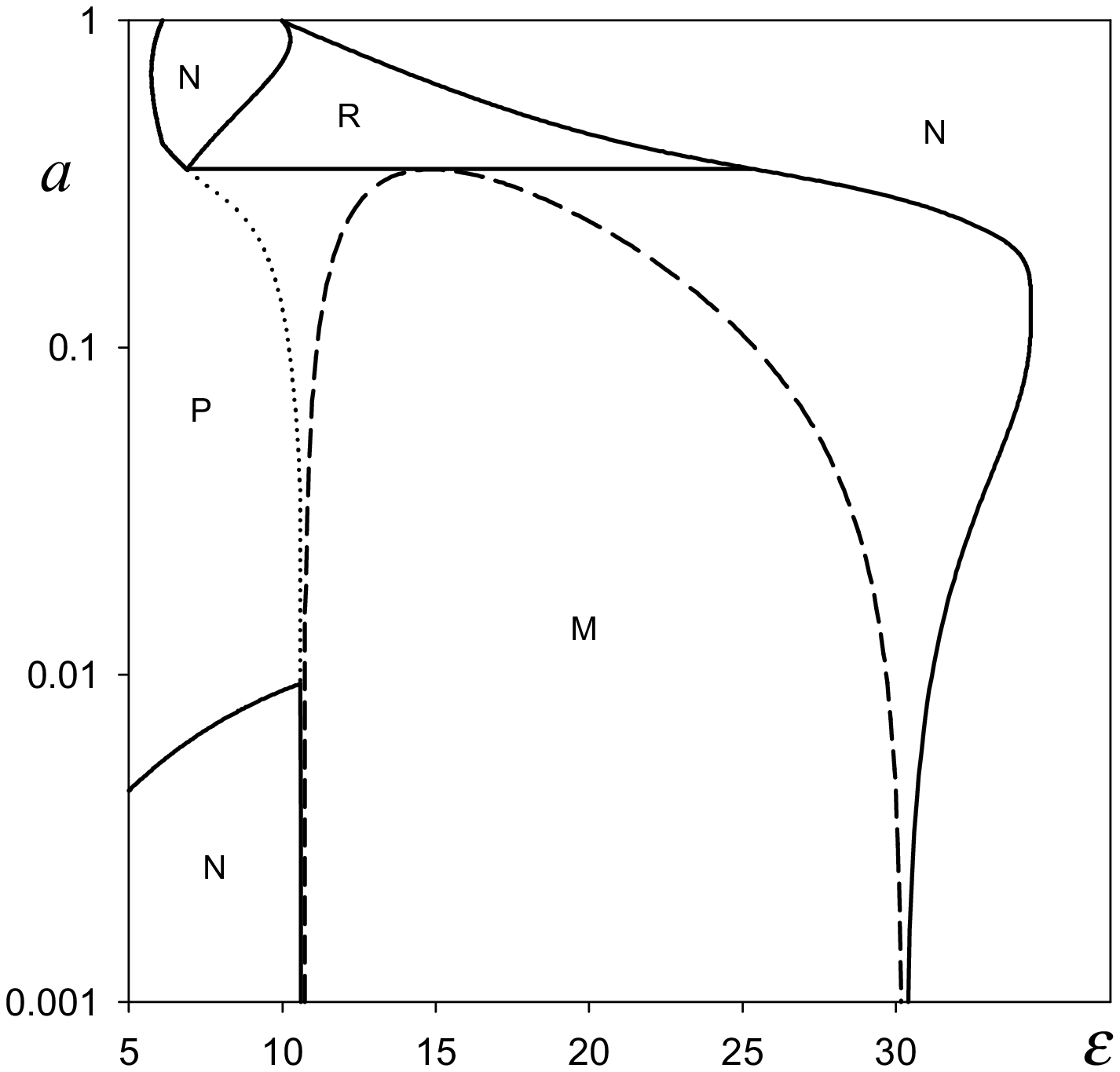}
\caption{Phase diagram in $(\ve, a)$ plane at $\alpha=0$, $\sigma^2_a=4.84$,
$\sigma^2_m=0.01$, $\tau_m=0.01$, $\tau_c=2.5$, $D=1.0$.
Dashed curve relates to the limit $\tau_c\to 0$.
\label{a(e)}}
\end{figure}

To find relations between noise exponent $a$ and the control
parameter $\ve$ we consider the phase diagram in $(\ve, a)$ plane.
It is appeared for noises with weak cross-correlation ($\tau_c\to
0$) the new phase arises only at small enough values of $a$ which
define the power of the multiplicative noise (see dashed curve in
Figure \ref{a(e)}).
In other words, considering the class of systems with both additive
and multiplicative noises, one should mean that ordering processes
are possible in the case of weak cross--correlation only if the
multiplicative noise has a weak power. For the systems with $a\to
1$ weak cross--correlation can not induce new phase formation.
According to solid--dotted curves in Figure \ref{a(e)}, an increase in
$\tau_c$ leads to appearance of the $\ve$ small valued domain
where the reorientation transition takes place with $a$--growth.
Besides, domains of both positive and negative order parameters,
being reoriented, join with metastable phase region at small
values of index $a$. However, an increase in $a$ leads to the reentrant phase
transition for long range cross--correlations. At
small and moderate values of the noise exponent $a$ the system behavior is
inherent in the hysteresis loop formation.

\begin{figure}[t]
\centering
\includegraphics[width=70mm]{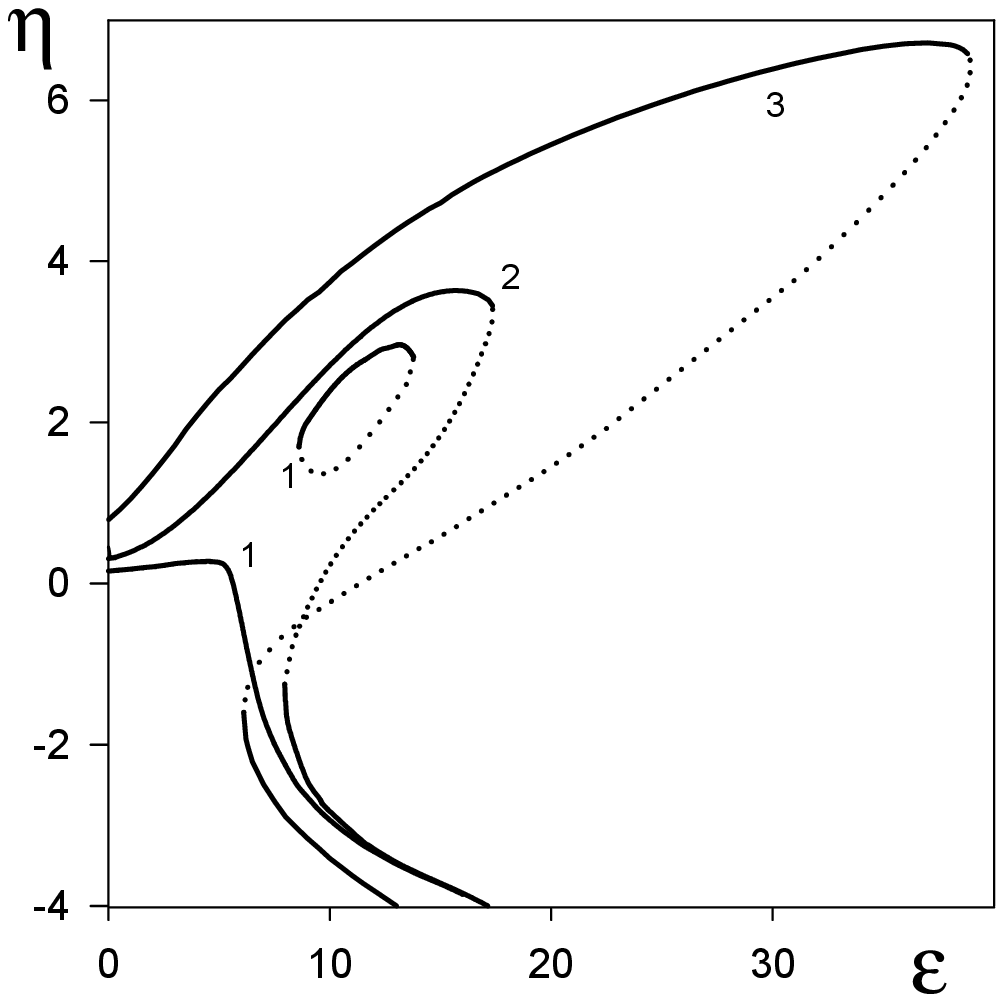}
\caption{The order parameter $\eta$ vs. the control parameter $\ve$ at  $\tau_m=0.01$,
$\sigma^2_a=4.84$, $\sigma^2_m=0.01$, $D=1.0$, $a=0.8$: curve 1 ---
$\alpha=0.2$, $\tau_c=2.5$; curve 2 --- $\alpha=0.2$, $\tau_c=5.0$; curve 3
---  $\alpha=0.7$, $\tau_c=5.0$. \label{eta(e)_alpha}}
\end{figure}

\begin{figure}[t]
\centering
a\\ \includegraphics[width=80mm]{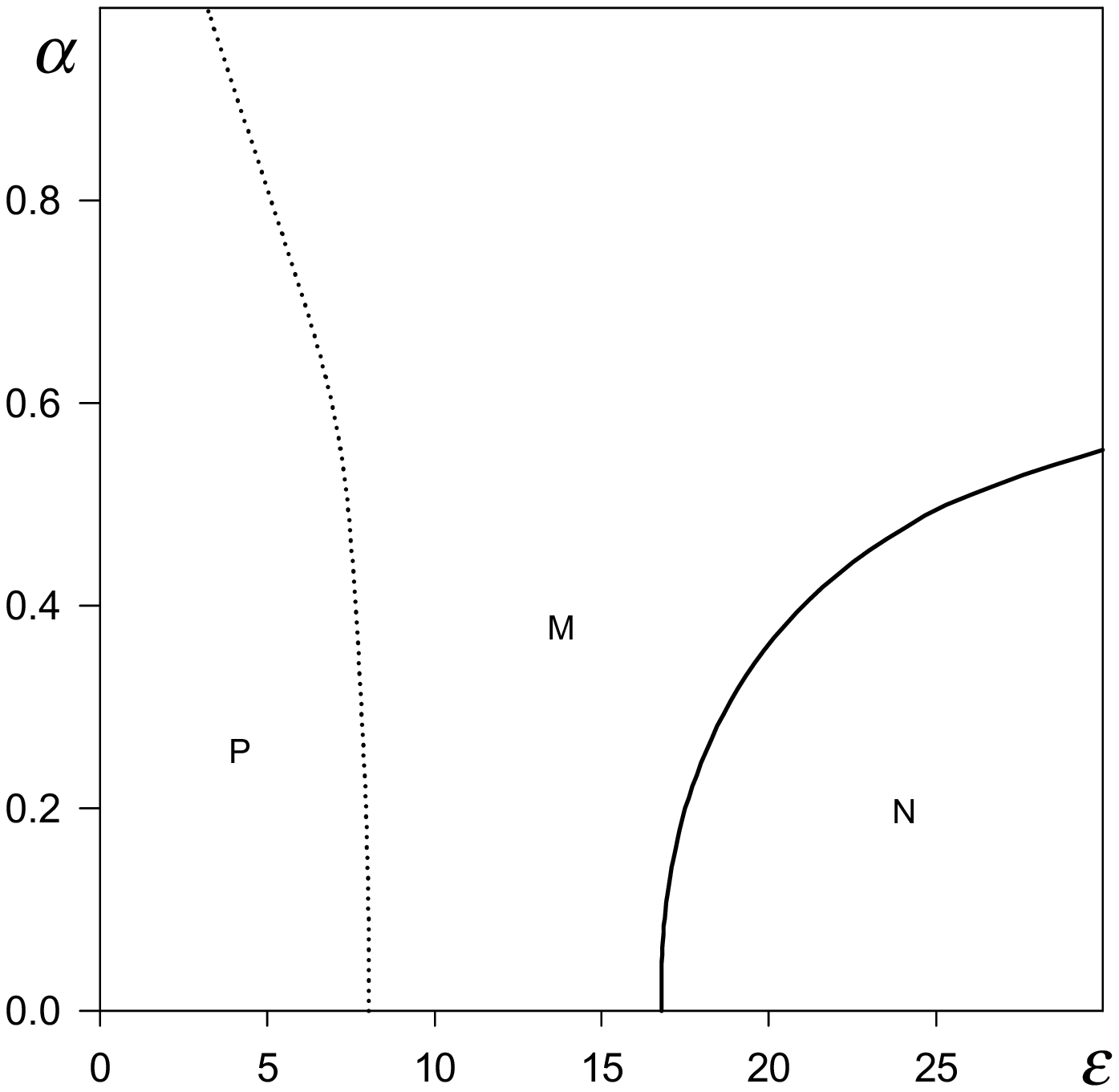}\\
b\\ \includegraphics[width=80mm]{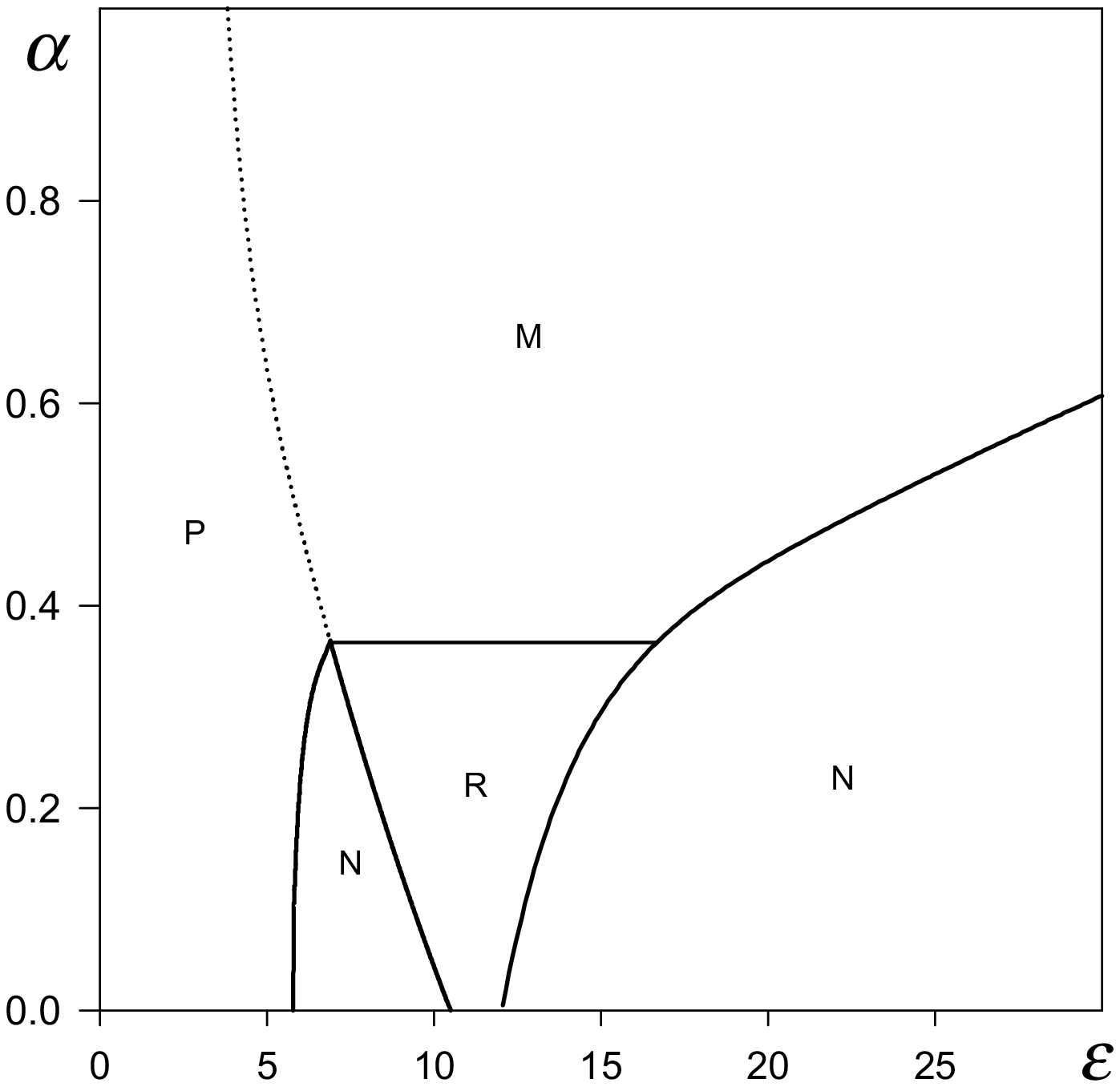}
\caption{Phase diagram $(\ve, \alpha)$ plane at
$\tau_m=0.01$, $\sigma^2_a=4.84$, $\sigma^2_m=0.01$, $D=1.0$, $a=0.8$:
a) $\tau_c=2.5$; b) $\tau_c=5.0$.\label{alpha(e)}}
\end{figure}

The influence of the damping exponent $\alpha\ne 0$
on the breaking symmetry picture is shown in Figure \ref{eta(e)_alpha}.
It is appeared an increase in $\alpha$ transforms the $\eta(\ve)$
dependence in a manner similar to the influence of the
cross--correlation time $\tau_c$. Indeed, passage from
two--connected $\eta(\ve)$ dependence related to curves 1 to
one--connected curves 2, 3 can be provided with both $\tau_c$
increase and $\alpha$ growth --- quite similarly to the
$\eta(\ve)$ dependencies variation in Figure \ref{eta(e)}. The
conclusion about similarity of the influences of the damping
exponent $\alpha\ne 0$ and the cross--correlation time $\tau_c$ is
confirmed with phase diagram in plane $(\ve,\alpha)$ which is
topologically identical to the same in Figure \ref{t(e)} at small
correlation times
(see Figure \ref{alpha(e)}a). According to the Figure \ref{alpha(e)}b
the cross--correlation time increase arrives at
the reentrant phase transition (within the domain R) due to appearance
of additional region N related to negative values of
the order parameter.

\begin{figure}[t]
\centering
a\\ \includegraphics[width=80mm]{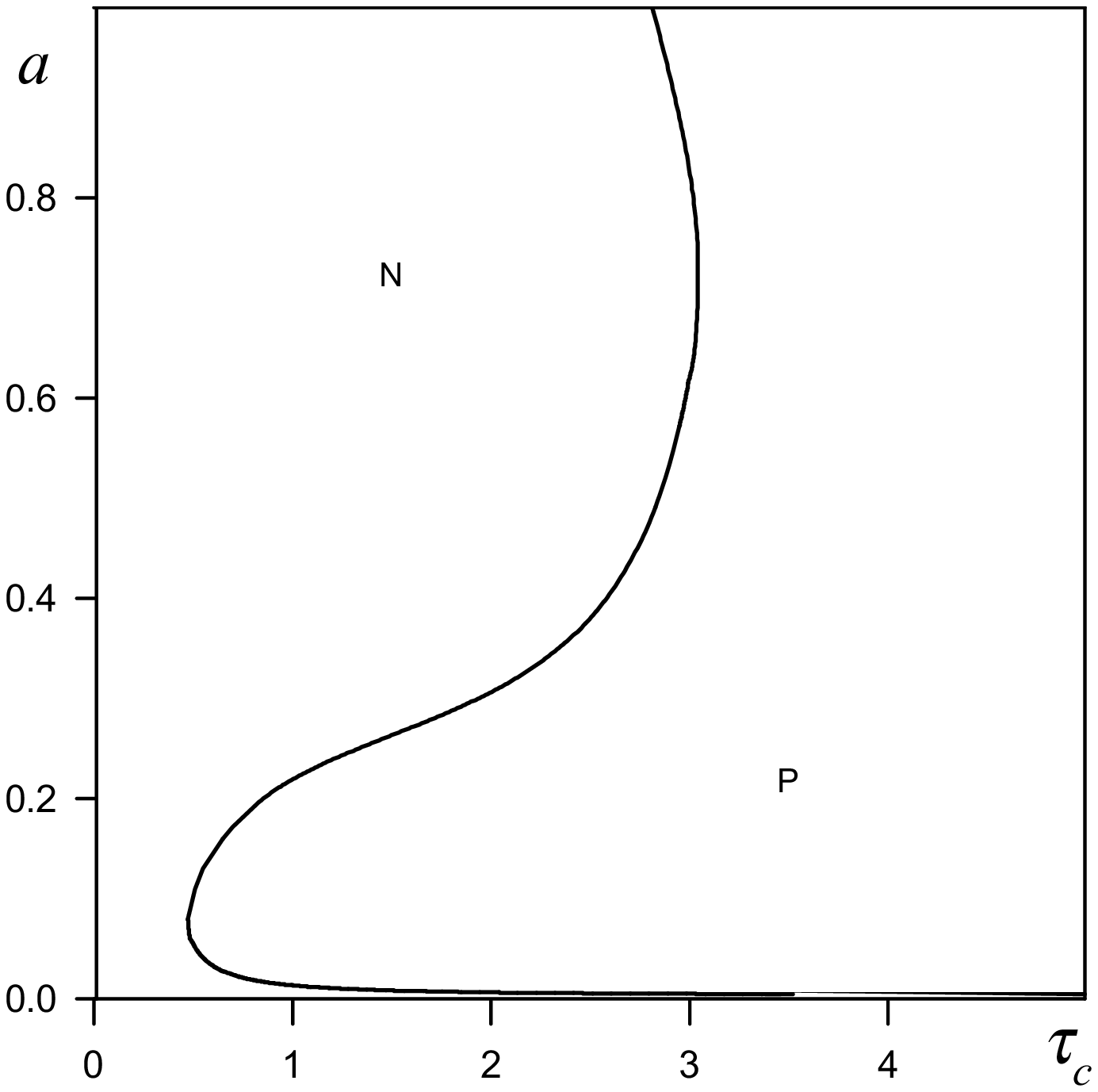}\\
b\\ \includegraphics[width=80mm]{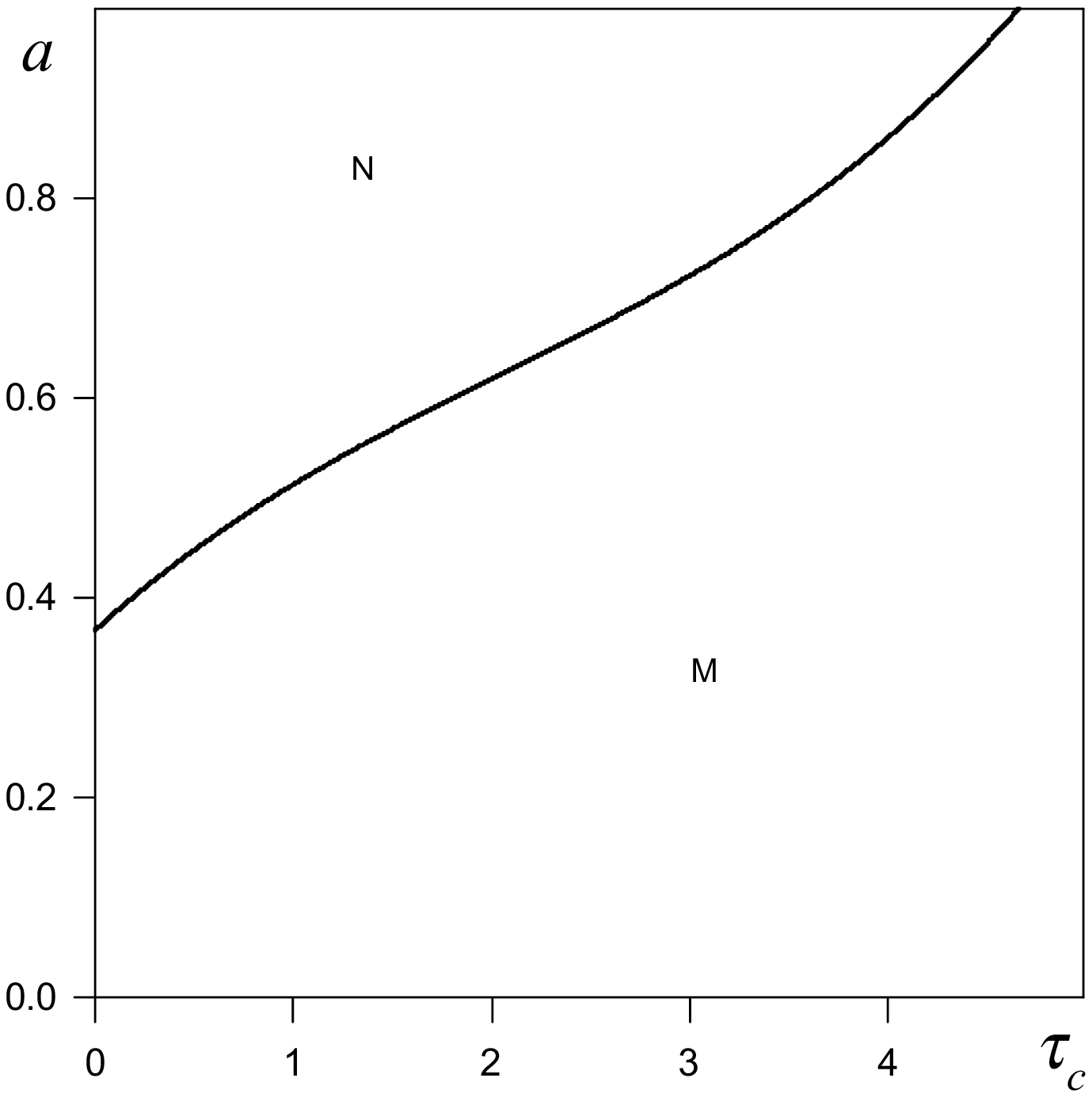}
\caption{Noise exponent $a$ vs. noise cross-correlation scale $\tau_c$ at
$\ve=6.5$ (a) and $\ve=15$ (b); other parameters are:  $\alpha=0$,
$\tau_m=0.01$, $\sigma^2_a=4.84$, $\sigma^2_m=0.01$, $D=1.0$.
\label{a(t)}}
\end{figure}

Finally, we set up the properties of the Langevin sources which allows us
to produce the ordering processes in the system. With this aim, we plot the
corresponding phase diagram in ($\tau_c, a$) plane (see Figure \ref{a(t)}).
Here, the system undergoes reorientation transition
related to the transforming the negative valued order parameter
into the positive one if $\tau_c$ increases.
On the other hand, at small $\ve$, increasing the exponent $a$
of multiplicative noise,
we can make the system to undergo a chain of phase transitions at which
the parameter $\eta$ changes the sign three times as maximum
(at $\ve=6.5$ and $\tau_c=2.85$ for example, see Figure \ref{a(t)}a).
The physical situation becomes more simple with an increase in $\ve$
(Figure \ref{a(t)}b).

\section{Discussion}

To understand main features of the system under consideration
we proceed from equation of effective motion
\begin{equation}
\dot x=\mathcal{D}_1(x)+\sqrt{\mathcal{D}_2(x)}\xi(t)
\label{ef}
\end{equation}
related to the Fokker--Planck equation (\ref{Fok_Plank}).
In difference of the initial noises $\zeta_\mu(t)$ in Eq.(\ref{1a})
effective noise $\xi(t)$ is of white--type:
$\langle\xi(t)\rangle=0$, $\langle\xi(t)\xi(0)\rangle=\delta(t)$.
Within the mean field approach, Eq.(\ref{ef}) takes the form
\begin{equation}
\dot\eta=\langle\mathcal{D}_1(x)\rangle\simeq
-\frac{{\partial}F}{{\partial}\eta},\quad
F\equiv\Delta F(\eta)-h\eta.
\label{eff}
\end{equation}
With accounting definition (\ref{D1a}), where the simplest
set of indexes $\alpha=0$, $a=1$ is chosen, a thermodynamic--type
potential $\Delta F(\eta)$ and
a field $h$ are defined by the following expressions:
\begin{equation}
\Delta F(\eta)=-\frac{\ve+\ve_m}{2}\eta^2+\frac{1}{4}\eta^4,\quad
\ve_m\equiv\tau_m\sigma_m^2;
\label{e1}
\end{equation}
\begin{equation}
h\equiv\tau_c\sigma_a\sigma_m.
\label{e2}
\end{equation}
Comparison of the first of definitions (\ref{e1})
with the bare potential (\ref{GL})
shows the multiplicative noise arrives at increase of the
control parameter $\ve$ due to addition $\ve_m$ whose magnitude is
proportional to the noise intensity $\sigma_m^2$ with the coefficient
$\tau_m$ being self--correlation time.
As a result, a growth of the multiplicative noise intensity in Figure
\ref{eta(e)_sigma} causes increasing the order parameter $\eta$ at
small magnitudes of the control parameter $\ve$.
A smearing of the related dependencies $\eta(\ve)$ at moderate values $\ve$
is caused by effective field $h$.

This field is inherent in the
cross--correlation effect fixed by the characteristic time
$\tau_c$ and intensities $\sigma_a$, $\sigma_m$ of both additive
and multiplicative noises. According to Eq.(\ref{eff}) the field
$h$ leads to deepening the right minimum of the thermodynamic
potential $F(\eta)$. If
cross--correlation effects are so slight that the condition
\begin{equation}
\tau_c<C~\frac{\left(\ve+\ve_m\right)^{3/2}}{\sigma_a\sigma_m},\quad
C\equiv\frac{2}{3^{3/2}}\simeq 0.385
\label{e3}
\end{equation}
is applied, the field $h$ is less than a critical value $h_c=
C(\ve+\ve_m)^{3/2}$ and the right minimum of the thermodynamic
potential $F(\eta)$ has a local character. It means the positive
order parameter  appears within a two--bounded interval
$\ve_c<\ve<\ve^c$ (see curves 2, 3 in Figure \ref{eta(e)}). With
strengthening cross--correlations, when the condition $h<h_c$
ceases to be valid, a barrier between right and left minima
disappears and a domain of definition of the positive order
parameter becomes bounded by the only upper boundary $\ve^c$
(curves 4, 5 in Figure \ref{eta(e)}).

With passage to the general case $a\ne 1$, the thermodynamic potential in
Eqs.(\ref{eff}), (\ref{e1}) takes the form
\begin{eqnarray}
&&F=\Delta F(\eta)-h~{\rm sign}(\eta)|\eta|^a,\label{e1bb}\\
&&\Delta F=-\left(\frac{\ve}{2}\eta^2+\frac{\ve_m}{2}\eta^{2a}\right)+
\frac{1}{4}\eta^4
\label{e1b}
\end{eqnarray}
that differs from the initial one by replacement
$\eta\to{\rm sign}(\eta)|\eta|^a$. As $a\leq 1$, this replacement derives to
more strong variations of the thermodynamic potential $V(\eta)$
within the actual domain $\eta<1$ that can arrive at the appearance
of local minimum at moderate values $\eta$. As a result,
a decrease of the index $a$ derives to metastable phase --- in perfect
accordance with Figure \ref{a(e)}.

To ascertain the effect of the index $\alpha$ we consider mean field approach
in the extreme case $\alpha=1$, $a=1$. Here, the thermodynamic potential
in the equation of motion (\ref{eff}) takes the form
\begin{equation}\label{e8}
F=F_{\lessgtr}\mp h\eta+\Delta F(\eta),\quad
F_{\lessgtr}\equiv\frac{1}{6}-\frac{\varepsilon+\varepsilon_m}{2}\pm\frac{4}{3}h
\end{equation}
where reference points $F_{\lessgtr}$ correspond to the domains
$\eta<-1$ and $\eta>1$, respectively, the field $h$ is determined
by Eq.(\ref{e2}) and the addition $\Delta
F(\eta)$ is defined by the expansion
\begin{equation}\label{e8a}
\Delta F\equiv\frac{A}{2}\eta^2+\frac{B}{3}\eta^3+\frac{C}{4}\eta^4
+\frac{E}{5}\eta^5+\frac{G}{6}\eta^6
\end{equation}
with the following coefficients:
\begin{equation}\label{e9}
\begin{split}
&A\equiv\mp(\varepsilon+\varepsilon_m)+ 2\sigma_a^{2},\quad
B\equiv(4\pm\tau_c)\sigma_a\sigma_m,\\
&C\equiv \pm \left[1+(\varepsilon+\varepsilon_m)\right]
+2(\sigma_m^{2}-\sigma_a^{2}),\\
&E\equiv -4\sigma_a\sigma_m,\quad
G\equiv \mp 1-2\sigma_m^{2};
\end{split}
\end{equation}
the upper and lower signs $\pm$ in the first Eq.(\ref{e8}) and
Eqs.(\ref{e9}) relate to the domains $|\eta|<1$ and
$|\eta|>1$, respectively. Comparing these equations with the
potential (\ref{e1}) addressed to the index $\alpha=0$ we convince
the dispersion of the damping coefficient (\ref{n3}) arrives at
transformation of the second order phase transition into the first
one --- as it follows from Figures \ref{eta(e)_alpha},
\ref{alpha(e)}.

\begin{figure}[h]
\includegraphics[width=80mm]{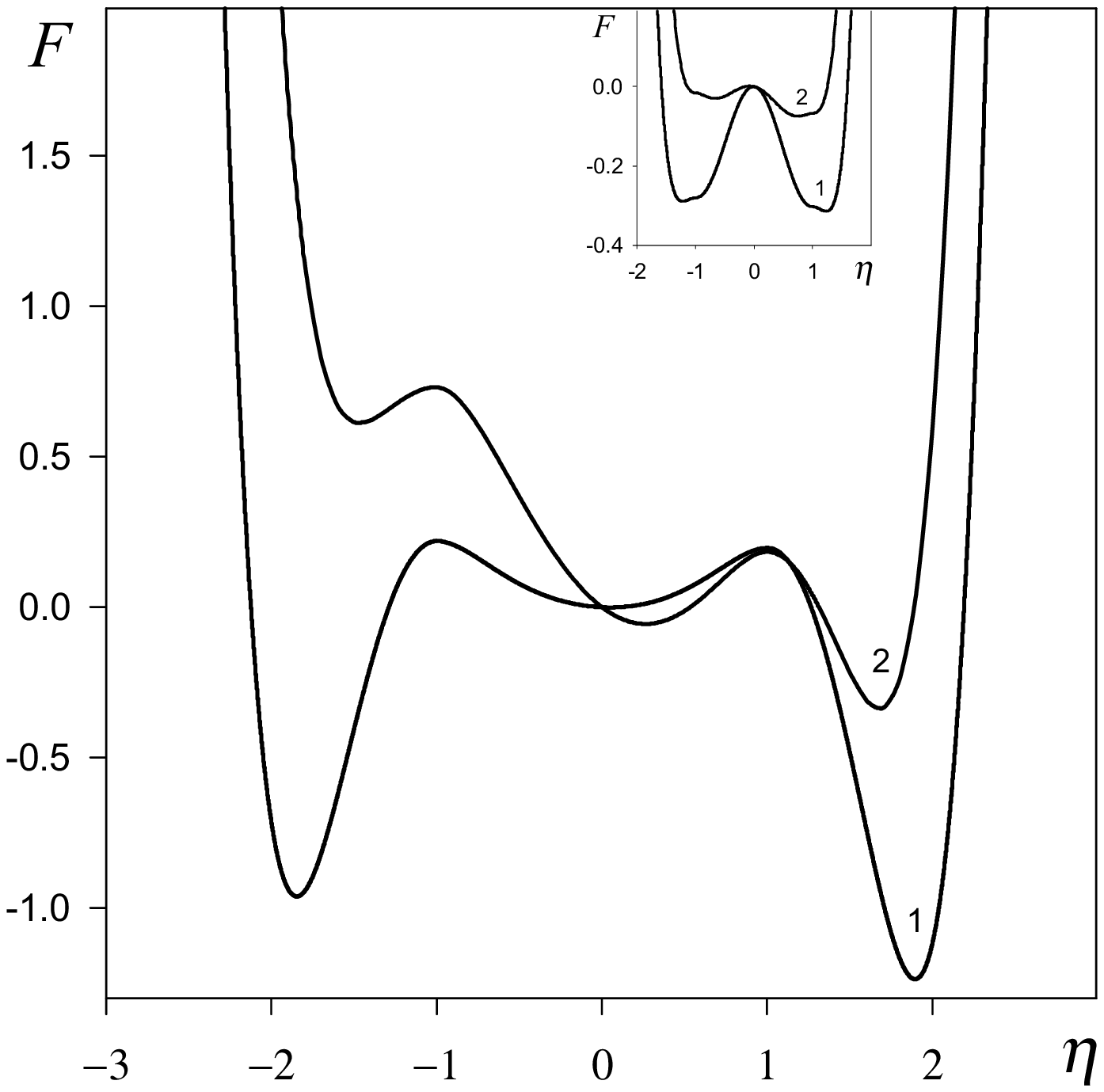}
\caption{The form of the thermodynamic potential $F$ given by
Eqs.(\ref{e8}) --- (\ref{e9}) (insertion shows the potential (\ref{ed8bab}),
curves 1 and 2 correspond to $D=0$ and $D=1$).}
\label{fig9}
\end{figure}

The form of the thermodynamic potential given by Eqs.(\ref{e8}) --- (\ref{e9})
is shown in Figure \ref{fig9} as a function of the order parameter.
It is seen this dependence has three well pronounced minima to be
inherent in the first order phase transition. Moreover, comparison
of the curves 1 and 2 shows the switching on the interaction field
(\ref{eta}) arrives at the gradient of the dependence $F(x)$
inducing the break symmetry. On the other hand, if we would like
to pass from the Ito calculus above used to the Stratonovich one,
we have to add the term $\frac{1}{2}\mathcal{D}_2^{\prime}(x)$ to
the drift coefficient $\mathcal{D}_1(x)$ in Eq. (\ref{eff})
\cite{VanKampen}. Then, the thermodynamic potential
$F\equiv-\int\mathcal{D}_1(\eta) {\rm d}\eta$ obtains the addition
$-\frac{1}{2}\mathcal{D}_2(\eta)$ to transform the potential given
by Eqs.(\ref{e8}) --- (\ref{e9}) to the form
\begin{equation}\label{ed8bab}
\begin{split}
&\widetilde F=\widetilde F_{\lessgtr}-{\tilde h}\eta+\Delta{\widetilde F}(\eta);\\
&\widetilde V_{\lessgtr}\equiv\frac{1}{6}-\frac{\varepsilon+
(\sigma_a^2+\tau_m\sigma_m^2)}{2}
\pm\frac{4}{3}\tau_c\sigma_a\sigma_m,\\
&{\tilde h}\equiv(1\pm\tau_c)\sigma_a\sigma_m,\\
&\Delta{\widetilde F}\equiv\frac{\tilde A}{2}\eta^2+\frac{\tilde B}{3}\eta^3
+\frac{\tilde C}{4}\eta^4
+\frac{\tilde E}{5}\eta^5+\frac{\tilde G}{6}\eta^6,\\
&\tilde A\equiv\mp\varepsilon+\left[4\sigma^2_a-(1\pm\tau_m)\sigma^2_m\right],\\
&\tilde B\equiv(10\pm\tau_c)\sigma_a\sigma_m,\\
&\tilde C\equiv\pm (1+\varepsilon)-\left[4\sigma^2_a-(6\pm\tau_m)\sigma^2_m
\right],\\
&\tilde E\equiv -9\sigma_a\sigma_m,\quad
\tilde G\equiv \mp 1-5\sigma_m^{2}
\end{split}
\end{equation}
where we take into account Eq.(\ref{D2a}) at $\alpha=1$, $a=1$.
Comparison of the corresponding dependencies $\widetilde F(\eta)$
shown in insertion of Figure \ref{fig9} with initial ones $F(\eta)$
depicted in the main panel shows the Stratonovich addition promotes
to transformation of the first order transition into the second one.

\begin{figure}[h]
\centering
a\\
\includegraphics[width=80mm]{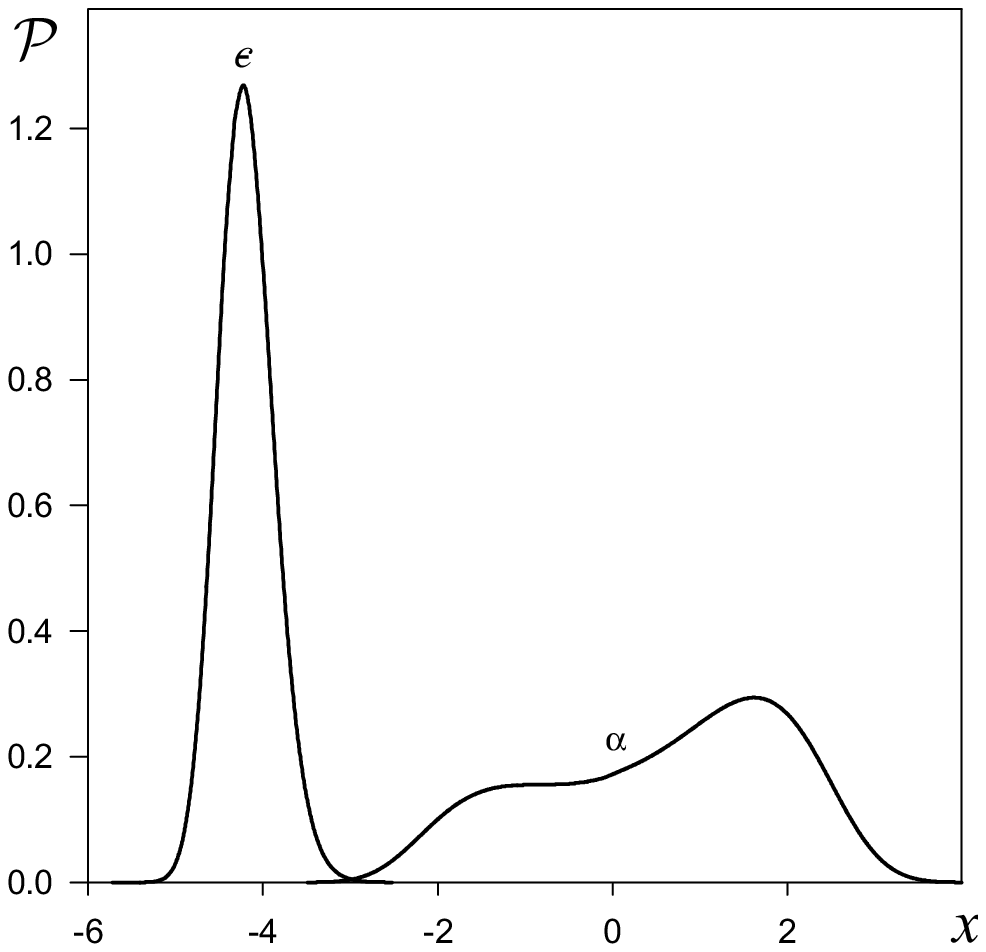}\\
b\\
\includegraphics[width=80mm]{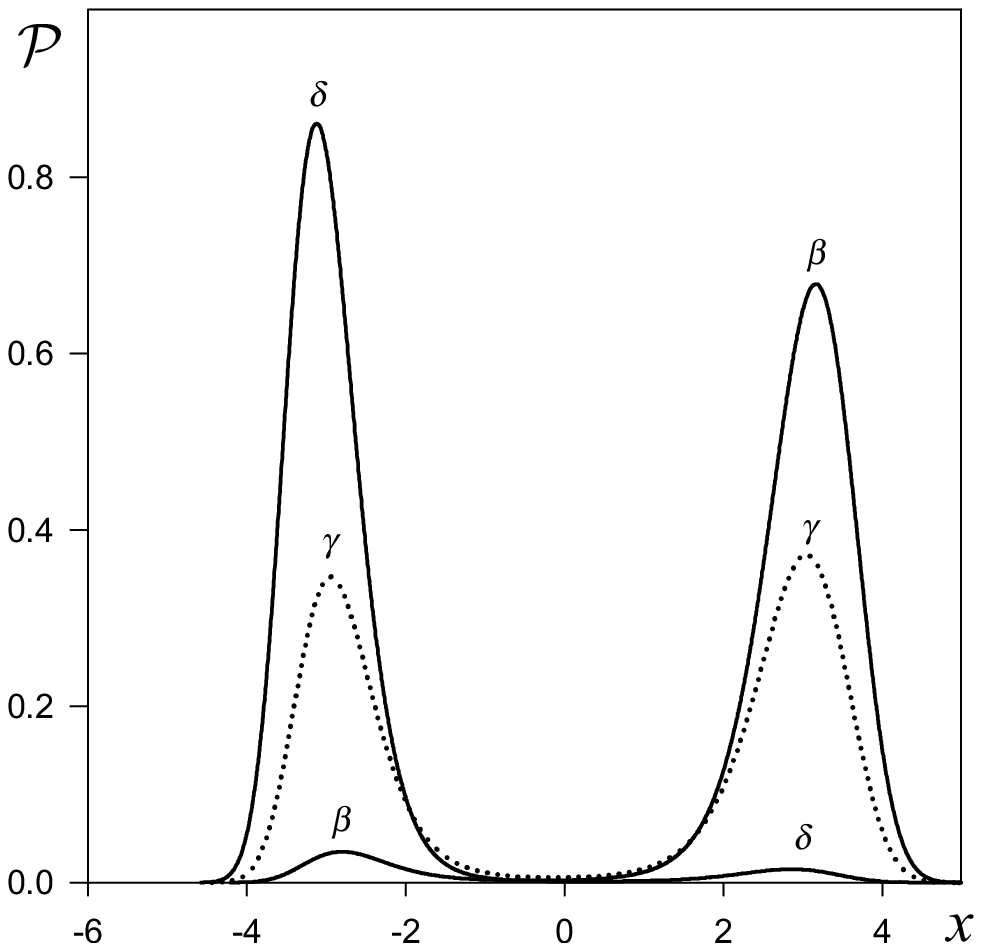}\\
\caption{Probability distributions addressed to different points
of the dependence $\eta(\ve)$ related to curve 1 in Figure
\ref{eta(e)_sigma}b: curves $\alpha$, $\epsilon$ in panel (a)
correspond to the values $\ve=3$, $\eta=0.56$ and $\ve=18$,
$\eta=-4.18$, respectively; in panel (b) the control parameter is
equal $\ve=12$ and curves $\beta$, $\gamma$, $\delta$ correspond
to different magnitudes of the order parameter $\eta=3.15$,
$\eta=1.14$ and $\eta=-3.32$.} \label{fig10}
\end{figure}

We proceed with consideration of the form of the probability distribution
function (\ref{pdf}) that is responsible for the reorientation transition
related to curve 1 in Figure \ref{eta(e)_sigma}b.
As shows comparison of the curves $\alpha$ and $\epsilon$ in the Figure
\ref{fig10}a,
positive magnitudes of the order parameter $\eta>0$ is related to
the distribution whose right maximum has a larger height
and is a wider than the left one (and vice versa at $\eta<0$).
Much more complicated picture takes place with growth of
the correlation time $\tau_c$ when strongly pronounced maximum
of the distribution (\ref{pdf}) is transformed from the left into the right one
by means of passage via the bimodal dependence
(see curves $\beta$, $\gamma$ and $\delta$ in Figure \ref{fig10}b).

\begin{figure}[h]
\includegraphics[width=80mm]{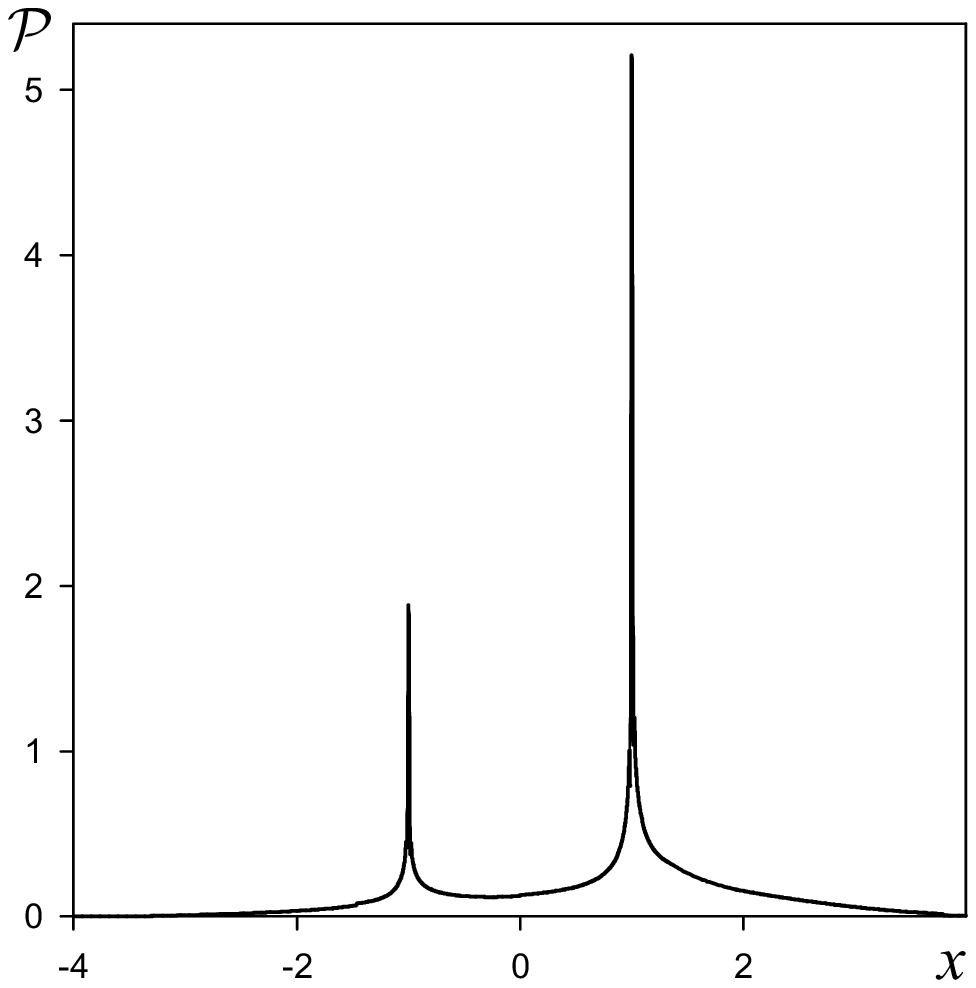}
\caption{Probability distribution related at $a=0.5$,
$\alpha=0.5$, $\ve=2$.}
\label{fig11}
\end{figure}

Above considered situation is picked out to address to the constant damping
coefficient (\ref{n3}) when the distribution (\ref{pdf})
has smooth form due to the index $\alpha=0$.
With passage to general case $\alpha\ne 0$ the dependence $\mathcal{P}_\eta(x)$
obtains the pair of the strong maxima in symmetrical points $x=\pm 1$
(see Figure \ref{fig11}).
The analytical form of these maxima follows from the estimations
\begin{equation}
\begin{split}
&\mathcal{D}_1(x)\simeq 2\alpha
\left(\sigma_a\pm\sigma_m\right)^2
x(x^2-1)^{2\alpha-1},\\
&\mathcal{D}_2(x)\simeq
\left(\sigma_a\pm\sigma_m\right)^2
|x^2-1|^{2\alpha}
\end{split}
\end{equation}
that are given by the dependencies (\ref{D1a}), (\ref{D2a})
near the points $x=\pm 1$.
As a result, we arrive at the integrable singularities
\begin{equation}
\mathcal{P}_\eta(x)\simeq\frac{\mathcal{Z}_\eta^{-1}}
{\left(\sigma_a\pm\sigma_m\right)^2|x^2-1|^\alpha}
\end{equation}
which have the form of the maxima
shown in Figure \ref{fig11}.
In contrary, near the point $x=0$ one has the estimations
\begin{equation}
\mathcal{D}_1(x)\simeq a\sigma_a\sigma_m\tau_c
|x|^{a-1}, \quad
\mathcal{D}_2(x)\simeq\sigma_a^2
\end{equation}
which arrive at the expression
\begin{equation}
\mathcal{P}_\eta(x)\simeq\frac{\mathcal{Z}_\eta^{-1}}{\sigma_a^2}
\exp\left(\frac{\sigma_m}{\sigma_a}\tau_c|x|^{a}\right)
\simeq\frac{\mathcal{Z}_\eta^{-1}}{\sigma_a^2}.
\end{equation}
Thus, the singularities of the drift coefficient $\mathcal{D}_1(x)$
at the point of origin has integrable character
to derive to the finite value of the probability distribution function.

\begin{figure}[h]
\includegraphics[width=80mm]{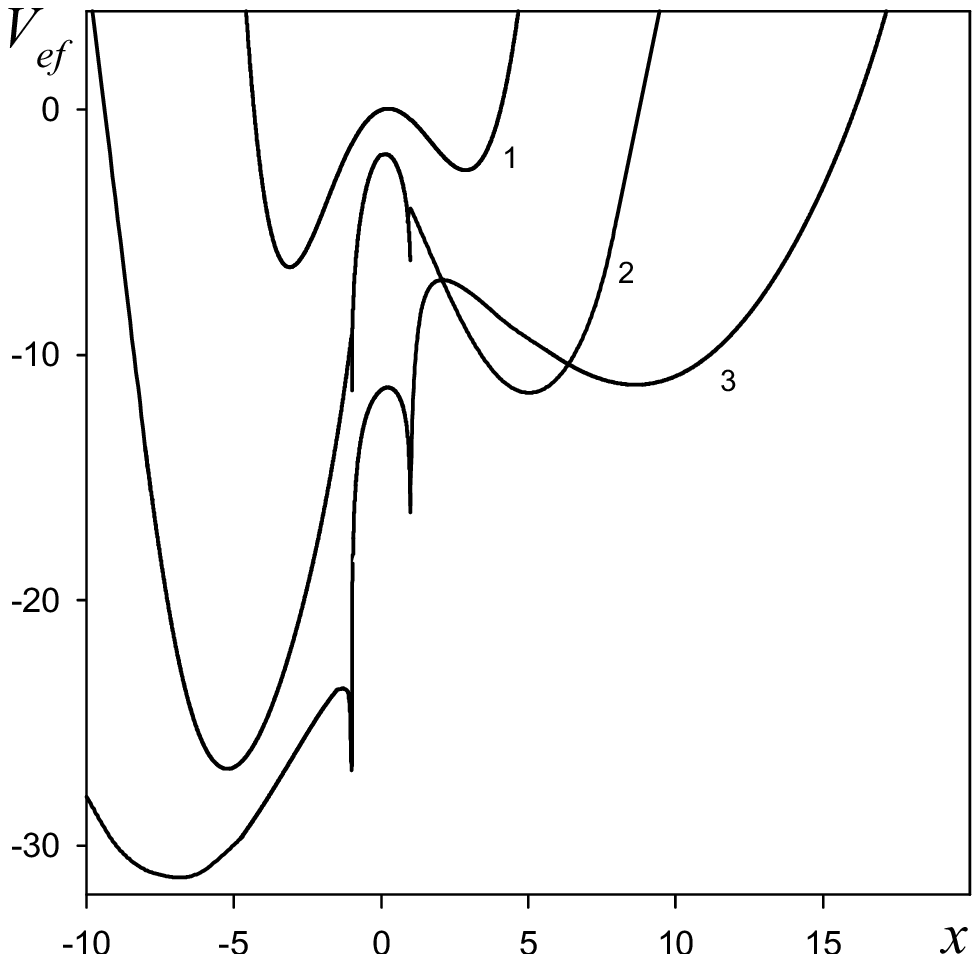}
\caption{The form of the effective potential (\ref{BGp}):
curves 1, 2 and 3 correspond to the cases
$a=1$, $\alpha=0$ ($\ve=10$, $\eta=-2.8$);
$a=0.5$, $\alpha=0.52$ ($\ve=10$, $\eta=-3.0$) and
$a=1$, $\alpha=1$ ($\ve=10$, $\eta=-4.88$), respectively.}
\label{fig12}
\end{figure}

Traditionally, one is taken to present the distribution function
(\ref{pdf}) in the Boltzmann--Gibbs exponential form
\begin{equation}\label{BG}
\mathcal{P}(x)\equiv
\exp\left\{-\frac{V_{ef}(x)}{\sigma_a^2}\right\}
\end{equation}
where an effective potential
\begin{equation}\label{BGp}
V_{ef}\equiv-\sigma_a^2\int\frac{\mathcal{D}_1(x)}{\mathcal{D}_2(x)}{\rm d}x
+\sigma_a^2\ln\mathcal{D}_2(x)
\end{equation}
is introduced to govern by the probability distribution in the
usual manner \cite{Horst}. Usage of the definitions (\ref{D1a})
and (\ref{D2a}) in the simplest case $a=1$, $\alpha=0$ derives to
explicit form of the potential (\ref{BGp}):
\begin{widetext}
\begin{equation}\label{BGpp}
\begin{split}
V_{ef}=&-\mathcal{H}x+\mathcal{V}(x);\qquad
\mathcal{H}\equiv D\eta+(\tau_c-2)\sigma_a\sigma_m,\quad
\mathcal{V}\equiv\frac{\mathcal A}{2}x^2+\frac{\mathcal B}{3}x^3+
\frac{\mathcal C}{4}x^4,\\
&\mathcal{A}\equiv(D-\varepsilon)-(2+\tau_m)\sigma^2_m
+2\frac{\sigma_m}{\sigma_a}(D\eta+\tau_c\sigma_a\sigma_m),\\
&\mathcal{B}\equiv 2\frac{\sigma_m}{\sigma_a}
\left[(\varepsilon-D)+(1+\tau_m)\sigma^2_m\right]
-3\left(\frac{\sigma_m}{\sigma_a}\right)^2
(D\eta+\tau_c\sigma_a\sigma_m),\\
&\mathcal{C}\equiv 1-\left(\frac{\sigma_m}{\sigma_a}\right)^2
\left[3(\varepsilon-D)+(2+3\tau_m)\sigma^2_m\right]
+4\left(\frac{\sigma_m}{\sigma_a}\right)^3
(D\eta+\tau_c\sigma_a\sigma_m)
\end{split}
\end{equation}
\end{widetext}
where we kept only the terms up to the fourth order of
the stochastic variable $x$. The form of the dependence
(\ref{BGp}) is depicted in Figure \ref{fig12} for different sets
of the indexes $a$ and $\alpha$. Comparison of the respective
curves confirms
the growth of the index $\alpha$ promotes to strengthening pair of the strong
minima at points $\eta=\pm 1$ of the bias dependence $V_{ef}(x)$.

\section{CONCLUSIONS}

In this paper we have considered the effect of the ordering of
stochastic system with two correlated noises. In so doing, we have
used a model of the system with Landau--like potential $V_0(x)$,
subject to both additive and multiplicative noises with amplitude
of the last in the form of the power--law function $|x|^a$,
$a\in[0,1]$ and affected by $x$--dependent damping with
coefficient $\gamma(x)=|x^2-1|^{-\alpha}$, $\alpha\in[0,1]$.
Within the framework of both the cumulant expansion method and
mean field theory, the stationary picture of the ordered states is
investigated in details. We have shown the fluctuation
cross--correlations can lead to the symmetry breaking of the
distribution function even in the case of the zero--dimensional
system. With introducing the spatial coupling, noise
cross--correlations can induce phase transitions where the order
parameter $\eta=\langle x\rangle$ varies discontinuously or in
reentrant manner. We have studied the specified interval of
magnitudes of system parameters where the ordered phase can be
formed. With this aim, principle phase diagrams are obtained to
illustrate the role of the multiplicative noise exponent $a$,
spectral characteristics of fluctuations (auto--correlation time
$\tau_m$ and cross--correlation time $\tau_c$), amplitudes of both
additive $\sigma_a$ and multiplicative $\sigma_m$ noises, exponent
$\alpha$ of the kinetic coefficient $\gamma(x)$, as well as
deterministic parameters being the dimensionless temperature
$\varepsilon$ and intensity of the spatial coupling $D$.

Above studied picture allows us to generalize the theory of the
phase transitions to the system with set of stochastic forces of
different nature. Basing on the mean field approach we have shown
the system can be described through the thermodynamic potential
$F(\eta)$ whose construction differs principally from the bare
potential $V_0(x)$: so, if the latter has the simplest
$x^4$--form, the former is shown to be of the $\eta^6$--form.
Coefficients of related expansion are obtained to define terms of
the even powers through the dimensionless temperature
$\varepsilon$, intensity of spatial coupling $D$,
auto--correlation time $\tau_m$ and intensities of both additive
$\sigma^2_a$ and multiplicative $\sigma^2_m$ noises;
terms of the odd powers are defined through the characteristics
$\tau_c$, $\sigma_a$ and $\sigma_m$ of the noise
cross--correlations, respectively. Thus, we can conclude the phase transition
tends to transform its character from continuous to discontinuous
due to the noise cross--correlation strengthening. This trend
displays more strongly with growth of the index $\alpha$ whose
value determines the dispersion of the damping coefficient
$\gamma(x)$. On the other hand, transition from the Ito calculus
to the Stratonovich one promotes to inverse transformation of the
discontinuous transition into the continuous one.

Obtained results can be applied to a consideration of the complex
systems which are far-off-equilibrium and hold  several collective
degrees of freedom. As shows consideration of three--dimensional
Lorenz--like system with noises being initially additive in nature,
usage of the saving principle reduces two of these noises to
multiplicative ones \cite{book1}. The physical reason of such a
picture is hierarchical subordination of different degrees of
freedom. According to our previous considerations \cite{rev}, \cite{book2}
a typical example of such type takes place in solid
state physics where a reentrant metastable phase can appear if the
matrix phase relates to random ensemble of defects of different
dimensions subject to the field of plastic flow (driven
dislocation--vacancy ensemble). Here, in the course of plastic
flow different defect structures alternate one another according
to picture of the first order phase transition. Moreover,
structural reorientation transitions take place where the sign of
the order parameter is related to the resulting direction of the
Burgers vectors of dislocation cluster. One more example of above
studied behavior gives reentrant glass transition in
colloid--polymer mixtures \cite{PRL}.

Note finally all presented results have been derived for the system with
nonconserved order parameter. The perspective of further exploration is to
investigate the system with conserved order parameter.

\section{ACKNOWLEDGEMENTS}

A. I. O. is gratefully acknowledged STCU, project 1976, for financial support.


\end{document}